\documentclass[prd,preprintnumbers,nofootinbib,preprint]{revtex4} 

\usepackage{epsfig,graphicx}
\usepackage{dcolumn}
\usepackage{bm}
\usepackage{latexsym}
\usepackage{amsfonts}
\usepackage{amssymb}
\usepackage{amsmath}

\begin{document}

\preprint{IPMU11-0189}

\title{Cosmological perturbations of self-accelerating universe in
nonlinear massive gravity}

\author{A. Emir G\"umr\"uk\c{c}\"uo\u{g}lu}
\email{emir.gumrukcuoglu@ipmu.jp}
\affiliation{
 IPMU, The University of Tokyo, Kashiwa, Chiba 277-8582, Japan}
\author{Chunshan Lin}
\email{chunshan.lin@ipmu.jp}
\affiliation{
 IPMU, The University of Tokyo, Kashiwa, Chiba 277-8582, Japan}
\author{Shinji Mukohyama}
\email{shinji.mukohyama@ipmu.jp}
\affiliation{
 IPMU, The University of Tokyo, Kashiwa, Chiba 277-8582, Japan}
\date{\today}

\begin{abstract}
 We study cosmological perturbations of self-accelerating universe
 solutions in the recently proposed nonlinear theory of massive gravity,
 with general matter content. While the broken diffeomorphism invariance 
 implies that there generically are $2$ tensor, $2$ vector and $2$ 
 scalar degrees of freedom in the gravity sector, we find that the
 scalar and vector degrees have vanishing kinetic terms and nonzero mass
 terms. Depending on their nonlinear behavior, this indicates either
 nondynamical nature of these degrees or strong couplings. Assuming the 
 former, we integrate out the $2$ vector and $2$ scalar degrees of
 freedom. We then find that in the scalar and vector sectors,
 gauge-invariant variables constructed from metric and matter
 perturbations have exactly the same quadratic action as in general
 relativity. The difference from general relativity arises only in the
 tensor sector, where the graviton mass modifies the dispersion relation
 of gravitational waves, with a time-dependent effective mass. This may
 lead to modification of stochastic gravitational wave spectrum. 
\end{abstract}

\maketitle

\section{Introduction}

The stability of general relativity (GR) predictions against small
graviton mass has been a persistent challenge of classical field
theory. The simplest ghost-free extension of GR with a linear mass term
\cite{Fierz:1939ix} suffers from the van Dam-Veltman-Zakharov
discontinuity, giving rise to different predictions for the classical
tests in the vanishing mass limit \cite{vanDam:1970vg,
Zakharov:1970cc}. Although this problem can be alleviated by nonlinear
terms \cite{Vainshtein:1972sx}, the cost is the emergence of the
Boulware-Deser (BD) ghost \cite{Boulware:1973my}, which is generically
unavoidable due to six degrees of freedom in the metric instead of the
five of the massive spin 2 field. 

Adopting an effective field theory approach in the decoupling limit, the
source of all the above issues can be traced back to the helicity 0 mode
of the graviton \cite{ArkaniHamed:2002sp}. In this perspective, an
analogue of the cancellation of the BD ghost in the linear theory by a
specific choice of the mass term can be performed, rendering the
nonlinear theory ghost-free up to quartic order by tuning the
coefficients \cite{Creminelli:2005qk}. 

More recently, a two parameter theory of nonlinear massive gravity
 was developed \cite{deRham:2010ik, deRham:2010kj}. In this
construction, terms to each order are chosen to remove the additional
degree of freedom (would-be BD ghost) at the decoupling limit and it has
the potential to be free of the BD ghost at fully nonlinear level
\cite{Hassan:2011hr, deRham:2011rn, deRham:2011qq}. The more general
construction with dynamical auxiliary metrics was also claimed to be
free from the BD ghost at fully nonlinear order \cite{Hassan:2011tf,
Hassan:2011ea}.

The massive extensions of GR are known to allow self-accelerating
solutions \cite{Salam:1976as, Damour:2002wu, deRham:2010tw,  
Koyama:2011xz, Nieuwenhuizen:2011sq, Koyama:2011yg, Chamseddine:2011bu,
D'Amico:2011jj, Gumrukcuoglu:2011ew, Koyama:2011wx, Comelli:2011wq,
Volkov:2011an, vonStrauss:2011mq, Comelli:2011zm,
Berezhiani:2011xx}. This provides an application opportunity for the
recent formulations of massive gravity, as an alternative approach to
account for the current accelerated expansion. On the other hand, to
model the accelerating universe in the scope of this infrared modified
gravity theory, a cosmological solution is necessary. In the
construction of \cite{deRham:2010ik, deRham:2010kj} with Minkowski
fiducial metric, flat Friedmann-Robertson-Walker (FRW) universe cannot
be realized \cite{D'Amico:2011jj}, although cosmological solutions with
negative spatial curvature exist \cite{Gumrukcuoglu:2011ew}. For more
general constructions with a nondynamical \cite{Tanahashi-progress}
and dynamical \cite{vonStrauss:2011mq, Comelli:2011zm} auxiliary
metrics, maximally symmetric FRW with either curvature may be allowed. 

Given the nontrivial cosmological solutions which self-accelerate, it is
thus a necessity to understand their properties against
perturbations. In the light of the construction which removes the
additional degree (would-be BD ghost), the expectation is to have five
degrees of freedom associated with the five polarizations of the massive
graviton. On the other hand, the absence of BD ghost does not guarantee
that the theory is safe; one still needs to determine the conditions
under which the helicity 0 and 1 modes, which are pure gauge in the
massless theory, are stable \cite{Higuchi:1986py}. Furthermore, these
additional degrees should not be in conflict with observations. For
instance, because of the emergence of a new scalar degree in the gravity
sector, the Newtonian potential may acquire modifications from couplings
between the matter sector and the helicity 0 graviton. If this is the
case, the parameters of the theory can be restricted by e.g. the solar
system tests. 

The primary goal of the present paper is to address these questions. We
consider the cosmological solutions in massive gravity as background and
in the presence of a generic matter content, we present a gauge
invariant formulation of perturbations. Although as we argued above, we
expect the gravity sector to contain $5$ dynamical degrees of freedom,
at the level of the quadratic action, we show that the helicity 0 and 1
graviton modes have vanishing kinetic terms but finite masses.\footnote{
A similar situation was noted in the decoupling limit
\cite{deRham:2010tw,Koyama:2011wx}.} As a result, in addition to the
matter perturbations, the only dynamical degrees of freedom are the two
tensor polarizations of gravity waves. 

This is exactly the same number of degrees of freedom as in GR. For the
action quadratic in perturbations around the cosmological backgrounds,
the mass terms of the potentially ghost-free construction turn out to be
completely decoupled from the standard part  (Einstein-Hilbert and
matter terms) for scalar and vector modes. Thus, these modes evolve
identically to their counterparts in GR, except for the additional
cosmological constant contributed by the graviton mass. The only non
trivial effect on the dynamics occurs in the tensor modes, which
acquires a time dependent mass term determined by the fiducial metric of
the theory. 

The paper is organized as follows. In Section II, we review the setup
with a fiducial metric that is Minkowski and summarize the only
cosmological solution \cite{Gumrukcuoglu:2011ew} allowed in this
case. In Section III, we give a detailed study of the complete quadratic
action for perturbations in the presence of a general fiducial metric of
the FRW form, with an ansatz for the physical background metric also of
FRW type with arbitrary spatial curvature. As an example, we specify to
perturbations around the open universe solution
\cite{Gumrukcuoglu:2011ew} driven by a single scalar field matter in
Section IV. We conclude with Section V, where we summarize our
results. The paper is supplemented by a number of Appendices, where the
details of calculations are presented.

\section{Open FRW solution with Minkowski fiducial metric}
\label{setup}

In this section, we review the open FRW universe solution
\cite{Gumrukcuoglu:2011ew} in nonlinear massive gravity 
\cite{deRham:2010kj} coupled to general matter content.

The covariant action for the gravity sector is constructed out of the
four dimensional metric $g_{\mu\nu}$ and the four scalar fields
$\phi^a\,(a=0,1,2,3)$ called {\it St\"uckelberg fields}. The action
respects the Poincare symmetry in the field space, i.e. invariance under
the constant shift of each of $\phi^a$ and the Lorentz transformation
mixing them: 
\begin{equation}
 \phi^a \to \phi^a + c^a, \quad \phi^a \to \Lambda^a_b\phi^b. 
  \label{eqn:Poincare-tr}
\end{equation}
The following line element in the field space is invariant under these
transformations. 
\begin{equation}
 \eta_{ab}d\phi^ad\phi^b = -(d\phi^0)^2 + \delta_{ij}d\phi^id\phi^j. 
  \label{eqn:modulispacemetric}
\end{equation}
Indeed, this is the unique geometrical quantity in the field space of
$\phi^a$. Thus the action can depend on $\phi^a$ only through the
spacetime tensor
\begin{equation}
 f_{\mu\nu} \equiv 
  \eta_{ab}\partial_{\mu}\phi^a\partial_{\nu}\phi^b.  
\end{equation}
In this language, general covariance is spontaneously broken by the
vacuum expectation value (vev) of $f_{\mu\nu}$. By assumption, matter 
fields propagate on the physical metric $g_{\mu\nu}$, but are not
coupled to $f_{\mu\nu}$ directly. The tensor $f_{\mu\nu}$, constructed
from the invariant line element in the field space, is often called a 
{\it fiducial metric}. On the other hand, the spacetime metric
$g_{\mu\nu}$, on which matter fields propagate, is often called a 
{\it physical metric}.

The gravity action is the sum of the Einstein-Hilbert action (with the
cosmological constant $\Lambda$) $I_{EH,\Lambda}$ for the physical
metric $g_{\mu\nu}$ and the graviton mass term $I_{mass}$ specified
below. Adding the matter action $I_{matter}$, the total action is 
\begin{equation}
 I = I_{EH,\Lambda}[g_{\mu\nu}] + I_{mass}[g_{\mu\nu},f_{\mu\nu}] 
  + I_{matter}[g_{\mu\nu},\sigma_I], 
\end{equation}
where 
\begin{eqnarray}
 I_{EH,\Lambda}[g_{\mu\nu}] & = & 
  \frac{M_{Pl}^2}{2}\int d^4x\sqrt{-g}(R-2\Lambda),
  \\
 I_{mass}[g_{\mu\nu},f_{\mu\nu}] & = & 
  M_{Pl}^2m_g^2\int d^4x\sqrt{-g}\,
  ( {\cal L}_2+\alpha_3{\cal L}_3+\alpha_4{\cal L}_4), 
  \label{eqn:Imass}
\end{eqnarray}
and $\{\sigma_I\}$ ($I=1,2,\cdots$) represent matter fields. Demanding
the absence of ghost at least in the decoupling
limit~\cite{deRham:2010kj}, each contribution in the mass term
$I_{mass}$ is constructed as 
\begin{eqnarray}
 {\cal L}_2 & = & \frac{1}{2}
  \left(\left[{\cal K}\right]^2-\left[{\cal K}^2\right]\right)\,, \nonumber\\
 {\cal L}_3 & = & \frac{1}{6}
  \left(\left[{\cal K}\right]^3-3\left[{\cal K}\right]\left[{\cal K}^2\right]+2\left[{\cal K}^3\right]\right), 
  \nonumber\\
 {\cal L}_4 & = & \frac{1}{24}
  \left(\left[{\cal K}\right]^4-6\left[{\cal K}\right]^2\left[{\cal K}^2\right]+3\left[{\cal K}^2\right]^2
   +8\left[{\cal K}\right]\left[{\cal K}^3\right]-6\left[{\cal K}^4\right]\right)\,,
\label{lag234}
\end{eqnarray}
where the square brackets denote trace operation and 
\begin{equation}
{\cal K}^\mu _\nu = \delta^\mu _\nu 
 - \left(\sqrt{g^{-1}f}\right)^{\mu}_{\ \nu}\,.
\label{Kdef}
\end{equation}
The square-root in this expression is the positive definite matrix
defined through 
\begin{equation}
 \left(\sqrt{g^{-1}f}\right)^{\mu}_{\ \rho}
  \left(\sqrt{g^{-1}f}\right)^{\rho}_{\ \nu}
  = f^{\mu}_{\ \nu}\ (\equiv g^{\mu\rho}f_{\rho\nu}).
  \label{eqn:square-root}
\end{equation}

As already stated above, a vev of the tensor $f_{\mu\nu}$ breaks general
covariance spontaneously. Thus, in order to find FRW cosmological
solutions in this theory, we should adopt an ansatz in which not only 
$g_{\mu\nu}$ but also $f_{\mu\nu}$ respects the symmetry of the FRW
universes~\cite{Gumrukcuoglu:2011ew}. Since the tensor $f_{\mu\nu}$ is
the pullback of the Minkowski metric in the field space to the physical
spacetime, construction of such an ansatz is equivalent to finding a
flat, closed, or open FRW coordinate system for the Minkowski line
element. It is well known that the Minkowski line element does not admit
a closed FRW chart but allows an open FRW chart. For this reason, in
order to find open FRW solutions \cite{Gumrukcuoglu:2011ew}, we first 
perform the field redefinition from $\phi^a$ to new fields $\varphi^a$
so that $f_{\mu\nu}$ written in terms of $\varphi^a$ manifestly has the
symmetry of open FRW universes as 
\begin{equation}
 f_{\mu\nu} = -n^2(\varphi^0)
  \partial_{\mu}\varphi^0\partial_{\nu}\varphi^0
  + \alpha^2(\varphi^0)
  \Omega_{ij}(\varphi^k)\partial_{\mu}\varphi^i\partial_{\nu}\varphi^j,
  \label{eqn:fmunu}
\end{equation}
where $i,j=1,2,3$, and 
\begin{equation}
\Omega_{ij}(\varphi^k) = \delta_{ij} + 
 \frac{K\delta_{il}\delta_{jm}\varphi^l\varphi^m}
 {1-K\delta_{lm}\varphi^l\varphi^m}
 \label{eqn:Omegaij}
\end{equation}
is the metric of the maximally symmetric space with the curvature
constant $K$ ($<0$). Concretely, this is achieved by 
\begin{equation}
 \phi^0  = 
  f(\varphi^0)\sqrt{1-K\delta_{ij}\varphi^i\varphi^j} \,,\qquad
 \phi^i = \sqrt{-K}f(\varphi^0)\varphi^i\,,
  \label{eqn:field-redefinition}
\end{equation}
and 
\begin{equation}
 n(\varphi^0) = |\dot{f}(\varphi^0)|, \qquad
  \alpha(\varphi^0) = \sqrt{-K}|f(\varphi^0)|,
  \label{eqn:n-alpha-Minkowski}
\end{equation}
where $f$ is a function to be determined and $\dot{f}$ represents its
derivative. We then adopt the ``unitary gauge'' 
\begin{equation}
 \varphi^0 = t, \quad \varphi^i = x^i,
  \label{eqn:unitary-gauge}
\end{equation} 
so that 
\begin{equation}
 f_{\mu\nu}dx^{\mu}dx^{\nu} = 
  - (\dot{f}(t))^2\,dt^2
  + |K|\,(f(t))^2\,\Omega_{ij}(x^k)dx^idx^j\,. 
\label{etaline}
\end{equation}
This is nothing but the Minkowski line element in the open chart. For
the physical metric, we adopt the open FRW ansatz 
\begin{equation}
ds^2 = -N(t)^2dt^2+a(t)^2\,\Omega_{ij}(x^k) dx^idx^j\,. 
 \label{eqn:FRWbackground}
\end{equation}
Hereafter, we assume that $N > 0$ and $a>0$, without loss of
generality.

The background action now yields, up to boundary terms, 
\begin{equation}
I = M_{Pl}^2 \int dt\,d^3x Na^3 \sqrt{\Omega}
 \left(L_{EH}[N,a]+m_g^2L_{mass}[N,a,f] \right)
 + I_{matter}[N,a,\sigma_I]\,,
 \label{totactin}
\end{equation}
consisting of the Einstein-Hilbert part
\begin{equation}
L_{EH} = \frac{3\,K}{a^2}-\frac{3\,\dot{a}^2}{a^2\,N^2}\,,
\end{equation}
and the contribution from the mass term 
\begin{eqnarray}
L_{mass} &=& \left(1-\frac{\sqrt{-K}|f|}{a}\right) 
\left[6+4\,\alpha_3+\alpha_4 - \frac{\sqrt{-K}|f|}{a} \left(3+5\,\alpha_3 +2\,\alpha_4\right) - \frac{K\,|f|^2}{a^2}\,\left(\alpha_3+\alpha_4\right)\right]
\nonumber\\
&& +{\mathrm sgn}(\dot{f}/f)\frac{|f|\,\dot{a}}{N\,a}\nonumber\\
 && \quad\times
 \left[3\left(3+3\,\alpha_3+\alpha_4\right) - \frac{3\,\sqrt{-K}|f|}{a} \left(1+2\,\alpha_3 +\alpha_4\right) - \frac{K\,|f|^2}{a^2}\,\left(\alpha_3+\alpha_4\right)\right]\,.
\end{eqnarray}
Hereafter, an overdot represents derivative w.r.t. the time $t$.

Varying the action (\ref{totactin}) with respect to $f$ yields the
following constraint 
\begin{eqnarray}
& & 
\left[H-{\mathrm sgn}(\dot{f}/f)\frac{\sqrt{-K}}{a} \right] 
\nonumber\\
& & \quad \times \left[3+3\,\alpha_3 +\alpha_4 -\frac{2\,\sqrt{-K}\,|f|}{a}\left(1+2\,\alpha_3+\alpha_4\right)-\frac{K\,|f|^2}{a^2}\,\left(\alpha_3+\alpha_4\right)\right]=0\,,
\label{eqn:constraintf}
\end{eqnarray}
where the Hubble expansion rate of the physical metric is defined as 
\begin{equation}
H \equiv \frac{\dot{a}}{N\,a}\,. \label{eqn:def-Hubble}
\end{equation}

Out of the three solutions of the constraint (\ref{eqn:constraintf}),
the trivial solution $\dot{a} = {\mathrm sgn}(\dot{f}/f)\sqrt{-K}\,N$
corresponds to the Minkowski spacetime in open chart. The remaining two
branches of solutions are given by \cite{Gumrukcuoglu:2011ew} 
\begin{equation}
 \alpha(t) = X_{\pm}a(t)\,,
\qquad
X_{\pm} \equiv \frac{1+2\,\alpha_3+\alpha_4
 \pm\sqrt{1+\alpha_3+\alpha_3^2-\alpha_4}}
 {\alpha_3+\alpha_4}\ (>0)\,,
\label{eq:fsolcosmo}
\end{equation}
and describe FRW cosmologies with $K<0$.\footnote{Note that $X_{\pm}$ are
positive by definition since $\alpha(t)>0$ and we assumed
$a(t)>0$. If we instead assumed $a(t)<0$ then the corresponding
solutions would be $\alpha(t)=-X_{\pm}a(t)$ with the same $X_{\pm}$ and
we would conclude $X_{\pm}>0$ again. The essential reason for the
positivity of $X_{\pm}$ is that the square-root in (\ref{Kdef}) is the
positive one.\label{footnote:Xpositive}} In the present paper we will
focus only on these nontrivial cosmological solutions.

Using the above constraint and varying the action (\ref{totactin}) with
respect to $N$ and $a$, we obtain the remaining background equations 
\begin{eqnarray}
 3\,H^2 +\frac{ 3\,K}{a^2} & = &
 \Lambda_\pm +\frac{1}{M_{Pl}^2}\rho\,,
\nonumber\\
-\frac{2\dot{H}}{N} +\frac{ 2\,K}{a^2} & = &
 \frac{1}{M_{Pl}^2}(\rho+P), 
\end{eqnarray}
where $\rho$ and $P$ are the energy density and the pressure of matter
fields calculated from $I_{matter}$, and  
\begin{equation}
\Lambda_\pm \equiv -\frac{m_g^2}{\left(\alpha_3+\alpha_4\right)^2}\left[\left(1+\alpha_3\right)\left(2+\alpha_3 +2\,\alpha_3^2-3\,\alpha_4\right) \pm 2\,\left(1+\alpha_3 +\alpha_3^2-\alpha_4\right)^{3/2}\right]\,.
\label{lambdapm}
\end{equation}
Thus, for the cosmological solutions (\ref{eq:fsolcosmo}), the 
contribution from the graviton mass term $I_{mass}$ at the background
level mimics a cosmological constant with the value $\Lambda_\pm$.

\begin{figure}[ht]
\begin{center}
\includegraphics[width=0.48\textwidth]{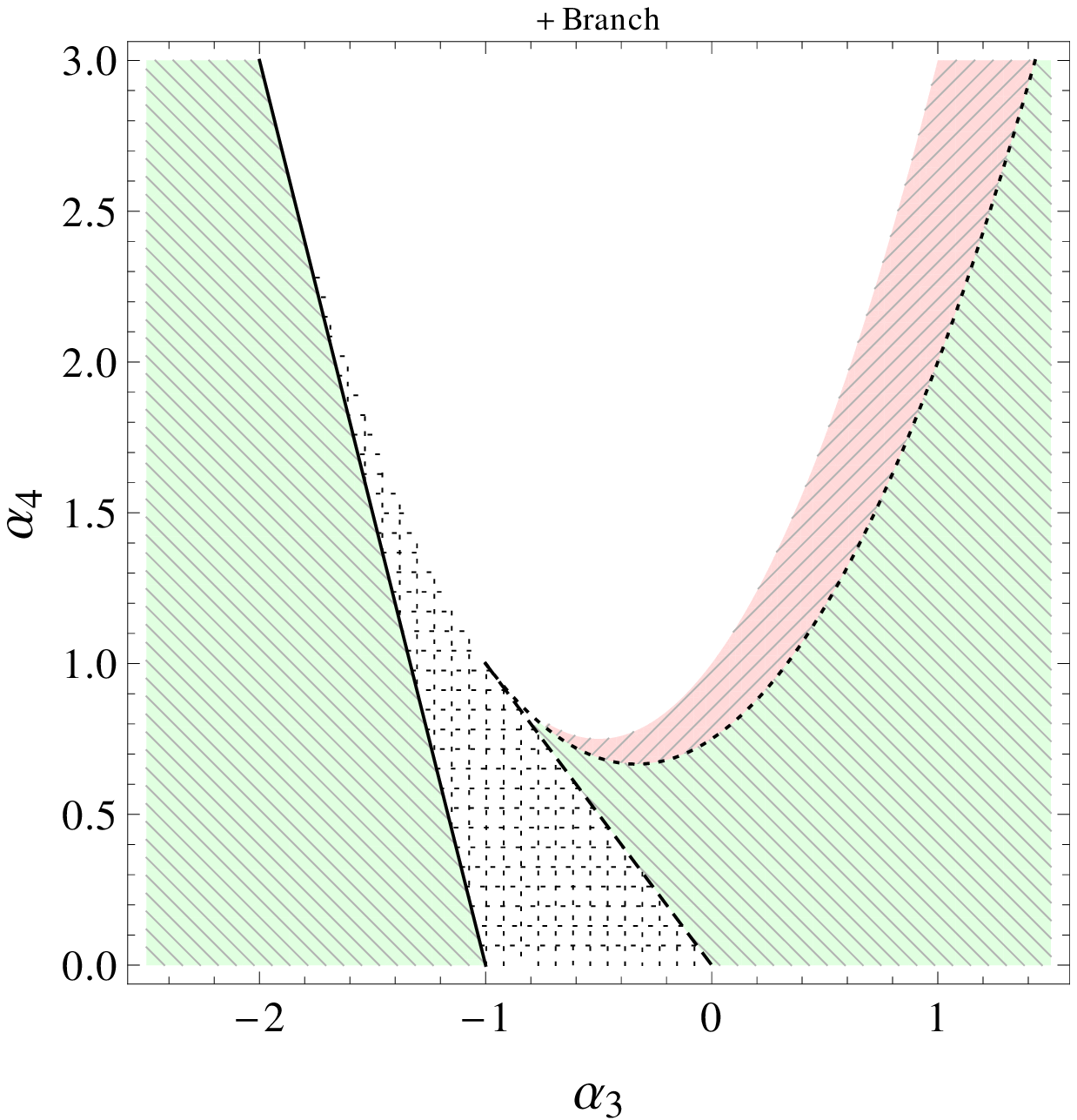}~~
\includegraphics[width=0.48\textwidth]{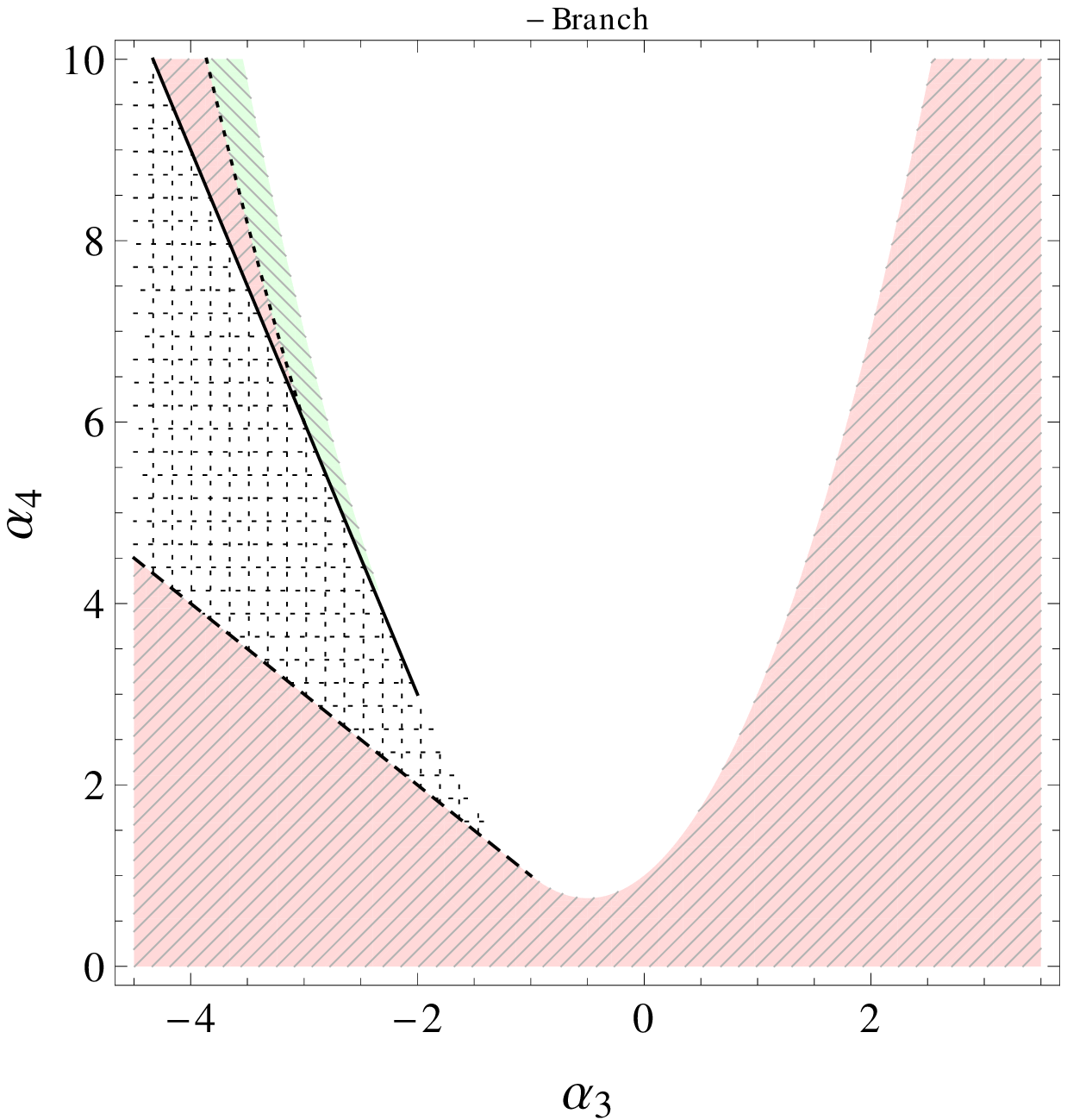}
\end{center}
\caption{Sign of the effective cosmological constant $\Lambda_\pm$ in
 the positive (left panel) and negative (right panel) branches. In the
 red (green) region with $+45^\circ$ ($-45^\circ$) lines, $\Lambda_{\pm}$ is
 positive (negative). The white region and the dotted squared region
 correspond to $1+\alpha_3+\alpha_3^2-\alpha_4<0$ and $X_{\pm}<0$,
 respectively, and are excluded since the cosmological solutions
 (\ref{eq:fsolcosmo}) do not exist there.  Along the dotted black line
 (defining the boundary between the red and green regions), 
 $\Lambda_\pm =0$ and the background solution reduces to the GR one. The
 solid line corresponds to $X_{\pm}=0$ and thus defines one of the
 boundaries between the allowed (red or green) and excluded (dotted
 squared) regions. Along the dashed line, both $X_{\pm}$ and
 $\Lambda_{\pm}$ diverge, and it defines another boundary between the
 allowed (red or green) and excluded (dotted squared) regions.}
 \label{fig:Lambdarange}
\end{figure}

For $\alpha_4=(3+2\,\alpha_3+3\,\alpha_3^2)/4$ and 
$\pm (1+\alpha_3)>0$, the effective cosmological constant
$\Lambda_\pm$ vanishes, and the background solution reduces to the
open FRW universe solution of GR. On the other hand, both $X_{\pm}$ and
$\Lambda_\pm$ diverge for $\alpha_4 =-\alpha_3$ and 
$\pm(1+\alpha_3)>0$. In Figure~\ref{fig:Lambdarange}, we show the sign
of $\Lambda_\pm$ in the $(\alpha_3, \alpha_4)$ space. Note that
$X_{\pm}$ are restricted to be positive by definition, as explained in
footnote \ref{footnote:Xpositive}. Except for the restriction due to the
positivity of $X_{\pm}$, these are in agreement with the analogous
region plots presented in
Ref.\cite{Koyama:2011wx}~\footnote{Substituting $\alpha_3 \to
3\,\alpha_3$, $\alpha_4 \to 12\,\alpha_4$, and switching the positive
and negative branch definitions, our expression (\ref{lambdapm})
recovers Eq.(6.6) of Ref.\cite{Koyama:2011wx}. However, note that
$f_{\mu\nu}$ in the solution of \cite{Koyama:2011wx} does not respect
the FRW symmetry. }. 

\section{Perturbations in general setup}
\label{sec:perturbations}

In this section we consider the graviton mass term
$I_{mass}[g_{\mu\nu},f_{\mu\nu}]$ defined by 
(\ref{eqn:Imass})-(\ref{Kdef}) and
(\ref{eqn:fmunu})-(\ref{eqn:Omegaij}), but with an arbitrary value of
$K$ and arbitrary functions $n(\varphi^0)$ and $\alpha(\varphi^0)$. We
shall develop a formalism to analyze perturbations of this generalized
system around flat ($K=0$), closed ($K>0$) and open ($K<0$) FRW 
universes. Cosmological implications of this type of generalized massive
gravity will be discussed in future publication.

For the background we adopt the physical metric of the FRW form
(\ref{eqn:FRWbackground}) with (\ref{eqn:unitary-gauge}), but with
general $K$, $n(\varphi^0)$ and $\alpha(\varphi^0)$. 
Without loss of generality, we assume that $N>0$, $n>0$, $a>0$ and
$\alpha>0$ at least in the vicinity of the time of interest, where $N$
and $a$ are the lapse function and the scale factor of the background
FRW physical metric. (Otherwise, we consider $|N|$, $|n|$, $|a|$ and
$|\alpha|$ and rename them as $N$, $n$, $a$ and $\alpha$.) As reviewed
in the previous section for open universes in the case of the Minkowski
fiducial metric and as shown in Appendix~\ref{app:gravitonmass} for
general cases with arbitrary $K$, $n(\varphi^0)$ and
$\alpha(\varphi^0)$, the background equation of motion for the
St\"uckelberg fields $\varphi^a$ has three branches of solutions. One of
them does not allow nontrivial cosmologies and thus is not of our
interest. The other two branches of solutions allow nontrivial
cosmologies and are given by (\ref{eq:fsolcosmo}) even for general $K$,
$n(\varphi^0)$ and $\alpha(\varphi^0)$. In this section we then consider
perturbations of the physical metric and the St\"uckelberg fields
around the FRW solutions in these nontrivial branches.

\subsection{Exponential map and Lie derivative}

Since the fiducial metric $f_{\mu\nu}$ is defined as in
(\ref{eqn:fmunu}) without referring to the physical metric $g_{\mu\nu}$,
let us begin with perturbations of the St\"uckelberg fields
$\varphi^a$. We define perturbations $\pi^a$ of $\varphi^a$ through the
so-called exponential map. Actually, since the action will be expanded
only up to the quadratic order, we can truncate the exponential map at
the second order. We thus define $\pi^a$ by
\begin{equation}
 \varphi^a = x^a + \pi^a + \frac{1}{2}\pi^b\partial_b\pi^a
  + O(\epsilon^3),\label{eqn:exponentialmap}
\end{equation}
or equivalently, 
\begin{equation}
 \pi^a = (\varphi^a-x^a)
  - \frac{1}{2}(\varphi^b-x^b)\partial_b(\varphi^a-x^a)
  + O(\epsilon^3). 
\end{equation}
Here, $\epsilon$ is a small number counting the order of perturbative
expansion: $\pi^a=O(\epsilon)$ and $\varphi^a-x^a=O(\epsilon)$. By
substituting the expansion (\ref{eqn:exponentialmap}) to the definition
of the fiducial metric 
\begin{equation}
 f_{\mu\nu} = \bar{f}_{ab}(\varphi^c)
  \partial_{\mu}\varphi^a\partial_{\nu}\varphi^b,
\end{equation}
where
\begin{equation}
 \bar{f}_{00}(\varphi^c) = -n^2(\varphi^0), \quad
  \bar{f}_{0i}(\varphi^c) = \bar{f}_{i0}(\varphi^c) = 0, \quad
  \bar{f}_{ij}(\varphi^c) = \alpha^2(\varphi^0)\Omega_{ij}(\varphi^k),
  \label{eqn:fbarab}
\end{equation}
we obtain
\begin{equation}
 f_{\mu\nu} = \bar{f}_{\mu\nu}(x^{\rho})
  + {\cal L}_{\pi}\bar{f}_{\mu\nu}(x^{\rho})
  + \frac{1}{2}\left({\cal L}_{\pi}\right)^2
  \bar{f}_{\mu\nu}(x^{\rho}) + O(\epsilon^3). 
  \label{eqn:fmunu-expansion}
\end{equation}
Here, ${\cal L}_{\pi}$ represents the Lie derivative along
$\pi^{\mu}$. Actually, this formula is not restricted to
(\ref{eqn:fbarab}) but holds for any $\bar{f}_{ab}(\varphi^c)$.

\subsection{St\"uckelberg fields and gauge invariant variables}
\label{subsec:gauge-invariant-variables}

We define perturbations $\phi$, $\beta_i$ and $h_{ij}$ of the physical 
metric by 
\begin{eqnarray}
 g_{00} & = & -N^2(t)\left[1+2\phi\right], \nonumber\\
 g_{0i} & = & N(t)a(t)\beta_i\nonumber\\
 g_{ij} & = & a^2(t)\left[\Omega_{ij}(x^k)+h_{ij}\right]. 
\end{eqnarray}
We suppose that $\phi,\beta_i,h_{ij}=O(\epsilon)$. 

Under the linear gauge transformation
\begin{equation}
 x^{\mu} \to x^{\mu} + \xi^{\mu},
  \quad (\xi^{\mu}=O(\epsilon))
  \label{eqn:gaugetr}
\end{equation}
each variable transforms as
\begin{eqnarray}
 \pi^0 & \to & \pi + \xi^0, \nonumber\\
 \pi_i & \to & \pi_i + \xi_i, \nonumber\\
 \phi & \to & \phi + \frac{1}{N}\partial_t(N\xi^0), \nonumber\\
 \beta_i & \to & 
  \beta_i - \frac{N}{a}D_i\xi^0 + \frac{a}{N}\dot{\xi}_i, \nonumber\\
 h_{ij} & \to & h_{ij} 
  + D_i\xi_j + D_j\xi_i + 2NH\xi^0\Omega_{ij},
\end{eqnarray}
where $H$ is the Hubble expansion rate as defined in
(\ref{eqn:def-Hubble}), 
\begin{equation}
 \pi_i \equiv \Omega_{ij}\pi^j, \quad
  \xi_i \equiv \Omega_{ij}\xi^j, 
\end{equation}
and $D_i$ is the spatial covariant derivative compatible with
$\Omega_{ij}$.

We then define gauge invariant variables 
\begin{eqnarray}
 \phi^{\pi} & \equiv & \phi - \frac{1}{N}\partial_t(N\pi^0), \nonumber\\
 \beta^{\pi}_i & \equiv & 
  \beta_i + \frac{N}{a}D_i\pi^0 - \frac{a}{N}\dot{\pi}_i, \nonumber\\
 h^{\pi}_{ij} & \equiv & h_{ij} 
  - D_i\pi_j - D_j\pi_i - 2NH\pi^0\Omega_{ij}. 
  \label{eqn:phipi-betapi-hpi}
\end{eqnarray}
For later convenience, let us decompose $\beta^{\pi}_i$ and
$h^{\pi}_{ij}$ as 
\begin{eqnarray}
 \beta^{\pi}_i & = & D_i\beta^{\pi} + S^{\pi}_i, \nonumber\\
 h^{\pi}_{ij} & = & 2\psi^{\pi}\Omega_{ij} + 
  \left(D_iD_j-\frac{1}{3}\Omega_{ij}\triangle\right)E^{\pi}
  + \frac{1}{2}(D_iF^{\pi}_j+D_jF^{\pi}_i) + \gamma_{ij}, 
  \label{eqn:decompose-hpi}
\end{eqnarray}
where $S^{\pi}_i$ and $F^{\pi}_i$ are transverse, and
$\gamma_{ij}$ is transverse and traceless:
\begin{equation}
 D^iS^{\pi}_i = D^iF^{\pi}_i = 0, \quad
  \quad D^i\gamma_{ij} = 0, \quad 
  \Omega^{ij}\gamma_{ij} = 0, 
\end{equation}
and $D^i\equiv\Omega^{ij}D_j$.

\subsection{Graviton mass term}

At the FRW background level, the graviton mass term acts as an effective
cosmological constant $\Lambda_{\pm}$ shown in (\ref{lambdapm}). The
proof of this statement is presented in Appendix~\ref{app:gravitonmass}
for arbitrary $K$, $n(\varphi^0)$ and $\alpha(\varphi^0)$. Thus,
calculations are expected to be simplified if we add 
$M_{Pl}^2\int d^4x\sqrt{-g}\Lambda_{\pm}$ to $I_{mass}$ before
performing perturbative expansion. For this reason, we define 
\begin{equation}
 \tilde{I}_{mass}[g_{\mu\nu},f_{\mu\nu}] \equiv 
  I_{mass}[g_{\mu\nu},f_{\mu\nu}]
  +M_{Pl}^2\int d^4x\sqrt{-g}\, \Lambda_{\pm},
  \label{eqn:def-Itildemass}
\end{equation}
and expand it instead of $I_{mass}$ itself.

As shown explicitly in Appendix~\ref{app:gravitonmass}, upon using the
background equation of motion for the St\"uckelberg fields but without
using the background equation of motion for the physical metric, the
graviton mass term can be expanded up to the quadratic order as 
\begin{eqnarray}
 \tilde{I}_{mass} & = &
  \tilde{I}_{mass}^{(0)} + \tilde{I}_{mass}^{(2)}[h^{\pi}_{ij}]
  + O(\epsilon^3), 
  \nonumber\\
 \tilde{I}_{mass}^{(2)}[h^{\pi}_{ij}] & = & 
  \frac{M_{Pl}^2}{8}\int d^4x Na^3\sqrt{\Omega}\, M_{GW}^2
  \left[ (h^{\pi})^2-h_{\pi}^{ij}h^{\pi}_{ij}\right], 
\end{eqnarray}
where the zero-th order part $\tilde{I}_{mass}^{(0)}$ is independent of
perturbations, 
\begin{eqnarray}
 M_{GW}^2 & \equiv & \pm (r-1)m_g^2\,
  X_{\pm}^2\sqrt{1+\alpha_3+\alpha_3^2-\alpha_4},\label{eqn:MGW2}\\
 r & \equiv & \frac{na}{N\alpha} = \frac{1}{X_{\pm}}\frac{H}{H_f}, 
  \quad
  H \equiv \frac{\dot{a}}{Na}, \quad
  H_f \equiv \frac{\dot{\alpha}}{n\alpha}, \nonumber
\end{eqnarray}
$X_{\pm}$ is given by (\ref{eq:fsolcosmo}), and
\begin{equation}
 h^{\pi} \equiv \Omega^{ij}h^{\pi}_{ij}, \quad
  h_{\pi}^{ij} \equiv \Omega^{ik}\Omega^{jl}h^{\pi}_{kl}.  
\end{equation}

With the decomposition of $h^{\pi}_{ij}$ in (\ref{eqn:decompose-hpi}), 
the quadratic mass term is expanded as
\begin{eqnarray}
 \tilde{I}_{mass}^{(2)}
  & = & M_{Pl}^2\int d^4x Na^3\sqrt{\Omega}\, M_{GW}^2
  \nonumber\\
 & & \times
  \left[3(\psi^{\pi})^2
   - \frac{1}{12}E^{\pi}\triangle(\triangle+3K)E^{\pi}
  + \frac{1}{16}F_{\pi}^i(\triangle+2K)F^{\pi}_i
  - \frac{1}{8}\gamma^{ij}\gamma_{ij} \right], 
\label{eqn:Imass2}
\end{eqnarray}
where 
\begin{equation}
 F_{\pi}^i \equiv \Omega^{ij}F^{\pi}_{j}, \quad
  \gamma^{ik} \equiv \Omega^{jl}\Omega^{jl}\gamma_{kl}.
\end{equation}

What is important here is that the quadratic part
$\tilde{I}_{mass}^{(2)}$ is gauge-invariant and depends only on
$h^{\pi}_{ij}$, or equivalently ($\psi^{\pi}$, $E^{\pi}$, $F^{\pi}_i$, 
$\gamma_{ij}$). In particular, it does not contribute to the equations
of motion for $\phi$ and $\beta_i$.

We note that $M_{GW}^2$ vanishes or diverges for some special values of
the parameters ($\alpha_3$, $\alpha_4$): 
\begin{eqnarray}
 \alpha_4 = -3\,(1+\alpha_3),\quad \pm(\alpha_3 +2)>0 
 & \Longrightarrow & {M}^2_{GW} =0\,,\nonumber\\
 \alpha_4 = 1+\alpha_3 +\alpha_3^2 \quad
  & \Longrightarrow & {M}^2_{GW} =0\,,
  \label{MGWzero}\\
 \alpha_4\to -\alpha_3, \quad
  \pm(1+\alpha_3)>0 \quad
  & \Longrightarrow & |{M}^2_{GW}| \to \infty, \nonumber
\end{eqnarray}
where the $\pm$ signs are for the $\pm$ branches, respectively. In the
following we suppose that the parameters ($\alpha_3$, $\alpha_4$) take
generic values away from the special values shown in (\ref{MGWzero}).

\subsection{Matter perturbations and gauge-invariant variables}

Let us divide matter fields $\sigma_I$ ($I=1,2,\cdots$) into the
background values $\sigma^{(0)}_I$ and perturbations as 
\begin{equation}
 \sigma_I = \sigma^{(0)}_I + \delta\sigma_I. \label{eqn:def-delta-sigma}
\end{equation}
We suppose that $\{\sigma_I\}$ forms a set of mutually independent
physical degrees of freedom. Otherwise, we consider a subset of the
original $\{\sigma_I\}$ consisting of independent physical degrees of
freedom and rename it as $\{\sigma_I\}$. We can construct
gauge-invariant variables $Q_I$ from $\delta\sigma_I$ and metric
perturbations, without referring to the St\"uckelberg fields.

For illustrative purpose let us decompose $\beta_i$, $h_{ij}$ and
$\xi_i$ as 
\begin{eqnarray}
\beta_i & = & D_i\beta + S_i, \nonumber\\
 h_{ij} & = & 2\psi\Omega_{ij} + 
  \left(D_iD_j-\frac{1}{3}\Omega_{ij}\triangle\right)E
  + \frac{1}{2}(D_iF_j+D_jF_i) + \gamma_{ij},
  \nonumber\\
 \xi_i & = & D_i\xi + \xi^T_i,
\end{eqnarray}
where $S_i$, $F_i$ and $\xi^T_i$ are transverse, and $\triangle$ is the 
Laplacian associated with $\Omega_{ij}$: 
\begin{equation}
 D^iS_i = D^iF_i = D^i\xi^T_i = 0, \quad
  \triangle \equiv D^iD_i.
\end{equation}
Under the gauge transformation (\ref{eqn:gaugetr}), each component of
the physical metric perturbation transforms as
\begin{eqnarray}
 \phi & \to & \phi + \frac{1}{N}\partial_t(N\xi^0), \nonumber\\
 \beta & \to & \beta -\frac{N}{a}\xi^0 + \frac{a}{N}\dot{\xi}, 
  \nonumber\\
 \psi & \to & \psi + NH\xi^0 + \frac{1}{3}\triangle\xi,
  \nonumber\\
 E & \to & E + 2\xi, \nonumber\\
 S_i & \to & S_i + \frac{a}{N}\dot{\xi}^T_i,
  \nonumber\\
 F_i & \to & F_i + 2\xi^T_i, \nonumber\\
 \gamma_{ij} & \to & \gamma_{ij}.
\end{eqnarray}
Noting that the vector $Z^{\mu}$ defined by 
\begin{eqnarray}
 Z^0 = -\frac{a}{N}\beta + \frac{a^2}{2N^2}\dot{E}, 
  \quad
  Z^i = \frac{1}{2}\Omega^{ij}(D_jE+F_j)
\end{eqnarray}
transforms as 
\begin{equation}
 Z^{\mu} \to Z^{\mu} + \xi^{\mu}\,,
\end{equation}
we can construct the following gauge-invariant variables out of
matter perturbations and physical metric perturbations:
\begin{eqnarray}
 Q_I & \equiv &
  \delta\sigma_I-{\cal L}_Z\sigma^{(0)}_I, \nonumber\\
 \Phi & \equiv & \phi - \frac{1}{N}\partial_t(NZ^0), \nonumber\\
 \Psi & \equiv & \psi - NHZ^0 - \frac{1}{6}\triangle E,
  \nonumber\\
 B_i & \equiv & S_i - \frac{a}{2N}\dot{F}_i\,,
\end{eqnarray}
and $\gamma_{ij}$ is gauge-invariant by itself. In the above, ${\cal L}_Z$ is
the Lie derivative along $Z^{\mu}$.

Those gauge-invariant variables defined here and in
Subsection~\ref{subsec:gauge-invariant-variables}, i.e. $\{Q_I$,
$\Phi$, $\Psi$, $B_i$, $\gamma_{ij}$, $\phi^{\pi}$, $\beta^{\pi}$,
$S^{\pi}_i$, $\psi^{\pi}$, $E^{\pi}$, $F^{\pi}_i\}$, are not
independent. Indeed, it is easy to show that 
\begin{eqnarray}
 \phi^{\pi} & = & \Phi + \frac{1}{N}\partial_t
  \left[\frac{1}{H}
  \left(\psi^{\pi}-\Psi-\frac{1}{6}\triangle E^{\pi}\right)
	 \right], 
  \nonumber\\
 \beta^{\pi} & = & 
  -\frac{1}{aH}\left(\psi^{\pi}-\Psi-\frac{1}{6}\triangle E^{\pi}\right)
  + \frac{a}{2N}\dot{E}^{\pi}, \nonumber\\
 S^{\pi}_i & = & B_i + \frac{a}{2N}\dot{F}^{\pi}_i. 
\end{eqnarray}
There are no more independent relations among gauge-invariant variables
defined here and in Subsection~\ref{subsec:gauge-invariant-variables}.\footnote{
See also the sentence just after (\ref{eqn:def-delta-sigma}).} Therefore,
we have the following set of independent gauge-invariant variables. 
\begin{equation}
 \{Q_I,\ \Phi,\ \Psi,\ B_i, \ \gamma_{ij}, \ 
  \psi^{\pi}, \ E^{\pi}, \ F^{\pi}_i\}. \label{eqn:independent-variables}
\end{equation}
Based on their origins, we can divide this set of independent
gauge-invariant variables into two categories as 
\begin{equation}
 \{Q_I,\ \Phi,\ \Psi,\ B_i, \ \gamma_{ij}\} 
  \quad \mbox{and} \quad
 \{\psi^{\pi}, \ E^{\pi}, \ F^{\pi}_i\}. 
\end{equation}
The first category consists of those gauge-invariant variables that
originate from the physical metric $g_{\mu\nu}$ and the matter fields
$\{\sigma_I\}$. Thus, those in the first category already exist in
GR coupled to the same matter content. On the other
hand, those in the second category are physical degrees of freedom
associated with the four St\"uckelberg fields $\varphi^a$.

\subsection{Structure of total quadratic action}

Let us now define 
\begin{equation}
 \tilde{I}[g_{\mu\nu},\sigma_I] \equiv 
  I_{EH,\tilde{\Lambda}}[g_{\mu\nu}]
  + I_{matter}[g_{\mu\nu},\sigma_I],\quad
  \tilde{\Lambda} \equiv \Lambda+\Lambda_{\pm},
\end{equation} 
so that 
\begin{equation}
 I = \tilde{I}[g_{\mu\nu},\sigma_I] 
  + \tilde{I}_{mass}[g_{\mu\nu},f_{\mu\nu}]. 
  \label{eqn:I=Itilde+Itildemass}
\end{equation}
Since $\tilde{I}_{mass}$ was already shown to be gauge-invariant up to
the quadratic order, (\ref{eqn:I=Itilde+Itildemass}) implies that
$\tilde{I}$ is also gauge-invariant up to that order. Thus the
quadratic part $\tilde{I}^{(2)}$ of $\tilde{I}$ can be written in terms
of gauge-invariant variables constructed solely from perturbations of the 
physical metric perturbations ($\phi$, $\beta_i$, $h_{ij}$) and matter
perturbations $\delta\sigma_I$, i.e. $\{Q_I$, $\Phi$, $\Psi$,
$B_i$, $\gamma_{ij}\}$.

Therefore, the total quadratic action has the following structure.
\begin{equation}
 I^{(2)} = 
  \tilde{I}^{(2)}[Q_I, \Phi, \Psi, B_i, \gamma_{ij}]
  + \tilde{I}^{(2)}_{mass}
  [\psi^{\pi},\ E^{\pi}, \ F^{\pi}_i, \gamma_{ij}],
\end{equation}
where the explicit form of $\tilde{I}^{(2)}_{mass}$ is shown in
(\ref{eqn:Imass2}). As already stated, gauge-invariant variables listed
in (\ref{eqn:independent-variables}) are independent from each other.

Note that $\psi^{\pi}$, $E^{\pi}$ and $F^{\pi}_i$ do not have kinetic
terms but have nonvanishing masses, provided that the parameters
($\alpha_3$, $\alpha_4$) take generic values away from the special
values shown in (\ref{MGWzero}). Thus, we can integrate them out: their
equations of motion lead to
\begin{equation}
 \psi^{\pi} = E^{\pi} = 0, \quad F^{\pi}_i = 0,
\end{equation}
and then
\begin{equation}
 I^{(2)} =   \tilde{I}^{(2)}[Q_I, \Phi, \Psi, B_i, \gamma_{ij}]
  - \frac{M_{Pl}^2}{8}\int d^4x Na^3\sqrt{\Omega}
  M_{GW}^2\gamma^{ij}\gamma_{ij}, 
\end{equation}
where $M_{GW}^2$ is given by (\ref{eqn:MGW2}). For scalar and vector
modes, this quadratic action is exactly the same as that in GR with the
matter content $\{\sigma_I\}$.

\subsection{Gravitational waves with time-dependent mass}

The total quadratic action for the tensor sector is 
\begin{equation}
 I^{(2)}_{tensor} = 
  \frac{M_{Pl}^2}{8}\int d^4x Na^3\sqrt{\Omega}
  \left[\frac{1}{N^2}\dot{\gamma}^{ij}\dot{\gamma}_{ij}
   + \frac{1}{a^2}\gamma^{ij}(\triangle-2K)\gamma_{ij}
   -M_{GW}^2\gamma^{ij}\gamma_{ij}\right],
  \label{eqn:I2tensor}
\end{equation}
provided that there is no tensor-type contribution from the quadratic
part of $I_{matter}$. In this way the dispersion relation of
gravitational waves is modified. The squared mass of gravitational waves 
$M_{GW}^2$ is given by (\ref{eqn:MGW2}) and is time-dependent.

If $M_{GW}^2$ is negative then long wavelength gravity waves exhibit
linear instability. For generic values of parameters ($\alpha_3$,
$\alpha_4$) away from the special values shown in (\ref{MGWzero}), we
see from the formula (\ref{eqn:MGW2}) 
that the sign of $M_{GW}^2$ is the same as the sign of the combination $\pm(r-1)m_g^2$, where $\pm$ signs correspond to $\pm$ branches, respectively.

\section{An example: scalar matter field and Minkowski fiducial}
\label{sec:example-scalar}

In the previous section we have analyzed quadratic action for
perturbations around nontrivial FRW backgrounds, with a general FRW
fiducial metric and a general matter content. For scalar and vector 
modes, we have shown that the quadratic action is exactly the same as
that in GR with the matter content. For tensor modes, on the other hand,
we have seen that gravitational waves obtain a time-dependent mass.

In this section, in order to illustrate these results a little more
explicitly, we shall consider a simple example consisting of the massive
gravity with Minkowski fiducial metric, coupled to a canonical scalar
matter field with potential $V$. The total action of this system is 
\begin{equation}
I = \int d^4x\sqrt{-g}
  \left[M_{Pl}^2\left(\frac{R}{2} 
  + m_g^2( {\cal L}_2+\alpha_3{\cal L}_3+\alpha_4{\cal L}_4) \right)
 -\frac{1}{2}\,\partial_\mu \sigma \partial^\mu \sigma- V(\sigma)\right]\,,
\end{equation}
where ${\cal L}_{1,2,3}$ are the graviton mass terms defined in
(\ref{lag234}). In the case of the Minkowski fiducial, the only
nontrivial FRW background is the open FRW solution found in
\cite{Gumrukcuoglu:2011ew} and reviewed in Sec.~\ref{setup}. Thus, in
this section the curvature constant $K$ is set to be negative and the
form of the fiducial metric is specified by
(\ref{eqn:fmunu})-(\ref{eqn:n-alpha-Minkowski}).

For the FRW background (\ref{etaline})-(\ref{eqn:FRWbackground}) with
$K<0$ and $\sigma=\sigma^{(0)}(t)$, the equations of motion read
\begin{eqnarray}
&&3\,H^2 +\frac{ 3\,K}{a^2} = \Lambda_\pm+\frac{1}{M_{Pl}^2}\left[\frac{(\dot{\sigma}^{(0)})^2}{2\,N^2}+V(\sigma^{(0)})\right]\,,
\nonumber\\
&&
-\frac{2\dot{H}}{N} +\frac{ 2\,K}{a^2} = \frac{(\dot{\sigma}^{(0)})^2}{M_{Pl}^2\,N^2}
\nonumber\\
&&
 \frac{1}{N}\partial_t\left(\frac{\dot{\sigma}^{(0)}}{N}\right)
 + \frac{3H}{N}\dot{\sigma}^{(0)} + V'(\sigma^{(0)}) = 0.
\label{bgeqs}
\end{eqnarray}

We now introduce perturbations to the metric $g_{\mu\nu}$, the four
St\"uckelberg fields $\varphi^a$ and the scalar matter field
$\sigma$. We will be developing the perturbation theory without
specifying a gauge and in the end we will switch to gauge invariant
perturbations, as we have already done in the previous section in a more 
general setup. The total quadratic action before switching to gauge
invariant perturbations is presented in Appendix~\ref{app:acttot}. We
then adopt the decomposition of the form 
\begin{eqnarray}
&&\qquad\quad \pi_i= D_i \pi + \pi_i^T \,, \qquad \beta_i = D_i\beta + S_i\,,
\nonumber\\
h_{ij} &=& 2\,\psi \,\Omega_{ij} + \left(D_i D_j -\frac{1}{3}\,\Omega_{ij} \triangle\right) E + \frac{1}{2}\left(D_i F_j + D_j F_i\right)+\gamma_{ij}\,,
\label{decomp}
\end{eqnarray}
where $\pi^T_i$, $S_i$ and $F_i$ are transverse: 
\begin{equation}
D^i \pi_i^T = D^i S_i = D^i F_i = 0\,.
\end{equation}

\subsection{Tensor sector}

We start by considering the tensor sector. We use the decomposition
(\ref{decomp}) in the total action (\ref{totact})-(\ref{totlag}),
keeping only the transverse traceless mode $\gamma_{ij}$. 
Since the matter sector has no tensor degrees, the action is the same as
the one given in (\ref{eqn:I2tensor}) with (\ref{eqn:MGW2}) and for the
Minkowski fiducial, we have 
\begin{equation}
 r = \frac{aH}{\sqrt{-K}}. 
\end{equation}
Notice that in the accelerating universe, $r$ and thus $M^2_{GW}$ grow 
at late time.

Switching to conformal time $d\eta \equiv N\,dt / a$ and defining the
canonical fields 
\begin{equation}
\bar{\gamma}_{ij} \equiv \frac{M_{Pl}\,a}{2}\,\gamma_{ij}\,,
\end{equation}
the tensor action takes the form
\begin{equation}
 I^{(2)}_{tensor} 
 = \frac{1}{2}\int d^3x\,d\eta\,\sqrt{\Omega}\,
 \left[\bar{\gamma}'_{ij}\bar{\gamma}^{\prime\,ij} 
  + \bar{\gamma}^{ij}
  \left(\triangle-2K+\frac{a''}{a}\right)\bar{\gamma}_{ij}
  - a^2M_{GW}^2\bar{\gamma}^{ij}\bar{\gamma}_{ij}
 \right]\,,
\end{equation}
where a prime denotes differentiation with respect to conformal
time. Next, we use harmonic expansion through 
\begin{equation}
\bar{\gamma}_{ij} = \int k^2 dk \,\bar{\gamma}_{\vec{k}} \,
 Y_{ij}(\vec{k},\vec{x})\,,
\end{equation}
where $Y_{ij}(\vec{k},\vec{x})$ is the tensor harmonic satisfying
\begin{equation}
\left(\triangle +\vec{k}^2\right)Y_{ij} = 0\,,\quad 
 D^i Y_{ij} = 0, \quad \Omega^{ij}Y_{ij} =0, \quad
 (\vec{k}^2 \equiv \Omega_{ij}\vec{k}^i\vec{k}^j)
\end{equation}
with $\vec{k}^2 \ge |K|$ taking continuous values. Suppressing the
momentum index, we obtain the equation of motion  
\begin{equation}
\bar{\gamma}'' + 
 \left(\vec{k}^2 - \frac{a''}{a} +2\,K + a^2\,M^2_{GW}\right)
 \bar{\gamma} =0\,.
\end{equation}
As we showed in the previous section for a generic setup, the tensor mode acquires a mass contribution whose time dependence is determined by the fiducial metric (which is Minkowski in the present example).

\subsection{Vector sector}

We now move on to the vector sector. Keeping only the transverse vector
modes in the decomposition (\ref{decomp}), the total action
(\ref{totact})-(\ref{totlag}) reduces to 
\begin{equation}
I^{(2)}_{vector}=\frac{M_{Pl}^2}{8} \int d^4xNa^3\sqrt{\Omega}\,
 {\cal L}_{vector}
\label{d2Ldef}
\end{equation}
with
\begin{eqnarray}
 {\cal L}_{vector} &=& \frac{1}{2\,N^2} \left(D^i \dot{F}^j D_i \dot{F}_j - 2\,K \dot{F}^i \dot{F}_i\right) 
-\frac{2}{N\,a} \left(D^i \dot{F}^j D_i S_j-2\,K \dot{F}^i S_i\right)
\nonumber\\
&&
-\frac{1}{2}\,M^2_{GW}\left(D^i F^j D_i F_j - 2\,K F^i F_i\right) 
+ \frac{2}{a^2}\,D^i S^j D_iS_j
-\frac{4\,K}{a^2}\,S_iS^i
\nonumber\\
&&
-2\,M_{GW}^2
\left(\triangle\pi^T_i+2\,K\,\pi^T_i\right)\left(F^i-\pi_T^i\right) 
\,,
\end{eqnarray}
where 
\begin{equation}
 S^i \equiv \Omega^{ij}S_j, \quad
  F^i \equiv \Omega^{ij}F_j, \quad
  \pi_T^i \equiv \Omega^{ij}\pi^T_j. 
\end{equation}
We then switch to gauge invariant perturbations, as defined in
(\ref{eqn:phipi-betapi-hpi})-(\ref{eqn:decompose-hpi}), and obtain 
\begin{eqnarray}
{\cal L}_{vector} &=& \frac{1}{2\,N^2} \left(D^i \dot{F}_{\pi}^j D_i \dot{F}^{\pi}_j-2\,K \dot{F}_{\pi}^i \dot{F}^{\pi}_i\right) 
-\frac{2}{N\,a} \left(D^i \dot{F}_{\pi}^j D_i S^{\pi}_j-2\,K \dot{F}_{\pi}^i S^{\pi}_i\right)
\nonumber\\
&&
-\frac{1}{2}\,M_{GW}^2 \left(D^i F_{\pi}^j D_i F^{\pi}_j - 2\,K F_{\pi}^i F^{\pi}_i\right) 
+ \frac{2}{a^2}\,D^i S_{\pi}^j D_iS^{\pi}_j -\frac{4\,K}{a^2}\,S_{\pi}^i S^{\pi}_i\,,
\end{eqnarray}
which is manifestly gauge invariant. Varying this action with
respect to $S^{\pi}_i$ yields an algebraic equation for $S^{\pi}_i$,
which can be solved by 
\begin{equation}
S^{\pi}_i = \frac{a}{2\,N}\,\dot{F}^{\pi}_i \,.
\end{equation}
Using this solution back in the action, we get 
\begin{equation}
I^{(2)}_{vector} =\frac{M_{Pl}^2}{16}\int d^4xNa^3\sqrt{\Omega}\,
 M^2_{GW}  \,F_{\pi}^i(\triangle+2K)F^{\pi}_i\,.
\end{equation}
This clearly shows that the kinetic term for vector perturbation
vanishes at quadratic order. Provided that the parameters ($\alpha_3$,
$\alpha_4$) take generic values away from the special values shown in
(\ref{MGWzero}), the equation of motion for $F^{\pi}_i$ leads to
$F^{\pi}_i=0$ and then $I^{(2)}_{vector}=0$.

\subsection{Scalar sector}

Finally, we consider the scalar perturbations. Using the decomposition
(\ref{decomp}) in the total action (\ref{totact})-(\ref{totlag}), we
obtain the action for the scalar sector as 
\begin{equation}
 I^{(2)}_{scalar} = 
  \frac{M_{Pl}^2}{2} \int d^4x Na^3\sqrt{\Omega}\,
  {\cal L}_{scalar}\,,
\end{equation}
with
\begin{eqnarray}
{\cal L}_{scalar}&=& 
\frac{1}{M_{Pl}^2}\left[\frac{1}{N^2}\,\delta\dot{\sigma}^2 - \frac{1}{a^2}\,D_i \delta \sigma D^i \delta \sigma-V''\,\delta\sigma^2 -\frac{2\,\dot{\sigma}}{N^2}\left(\phi-3\,\psi\right)\delta\dot{\sigma} - 2\,V'\,\left(\phi+3\,\psi\right)\delta\sigma\right]
\nonumber\\&&
+\frac{1}{N^2}\left(\frac{1}{6} (\triangle \dot{E})^2 - \frac{K}{2} D^i \dot{E} D_i \dot{E}-6\,\dot{\psi}^2 \right)
+\frac{12\, H}{N}\,\phi\,\dot{\psi}-\frac{12\,K}{a^2}\phi\,\psi 
\nonumber\\
&&- \frac{1}{6} \left( M^2_{GW}+\frac{K}{a^2}\right)(\triangle E)^2
+\frac{K}{2} M^2_{GW}D_iE\,D^iE
-\left(6\, H^2 -\frac{\dot{\sigma}^2}{M_{Pl}^2\,N^2}\right)\phi^2
\nonumber\\&&
-\frac{2\,K}{a^2}D_i\beta\,D^i\beta 
+\frac{2}{a\,N}\,\left(2\,\triangle\dot{\psi}-\frac{1}{3}\,\triangle^2 \dot{E}-K\,\triangle\dot{E} - 2\,N H\triangle\phi +\frac{\dot{\sigma}}{M_{Pl}^2}\,\triangle\delta\sigma
\right)\beta
\nonumber\\&&
-\frac{2}{a^2}\,\left(2\,\triangle\psi-\frac{1}{3}\triangle^2E - K\,\triangle E \right)\phi+6\,\left( M_{GW}^2 - \frac{K}{a^2}\right)\psi^2 
\nonumber\\&&
+\frac{1}{a^2}\,\left(2\,D_i\psi D^i\psi +\frac{2}{3}\,\triangle \psi \triangle E + \frac{1}{18}\,D_i \triangle E \,D^i \triangle E - 2\, K \,D_i\psi\,D^i E\right)
\nonumber\\&&
+2\, M^2_{GW}\,\left(K\left(D_i\pi\,D^i\pi-D_iED^i\pi\right)+\frac{1}{3}\triangle E\,\triangle \pi-2\,\psi\,\triangle\pi-6\, H\,N\,\psi\,\pi^0\right)
\nonumber\\&&
+2\, H\,N^2\, M^2_{GW}\,\left(\frac{2}{N}\,\triangle\pi+3\, H\pi^0\right)\pi^0\,.
\end{eqnarray}
Hereafter in this subsection, $\sigma$ represents the background value
$\sigma^{(0)}(t)$. Next, we switch to gauge invariant variables defined
in (\ref{eqn:phipi-betapi-hpi})-(\ref{eqn:decompose-hpi}) and 
carry out harmonic expansion. 
We then obtain the equations of motion for
the nondynamical degrees as
\begin{eqnarray}
\phi^{\pi} &=& \frac{1}{H}\,\left(\frac{K}{a}\, \beta^{\pi} +\frac{\dot{\psi}^{\pi}}{N}+\frac{k^2-3\,K}{6\,N}\,\dot{E}^{\pi}+\frac{\dot{\sigma}}{2\,M_{Pl}^2N}\,\delta\sigma^{\pi}\right)\,,\nonumber\\
\beta^{\pi} &=& -\frac{a}{k^2\,H}\,\left[\left(-3\,H^2+\frac{\dot{\sigma}^2}{2\,M_{Pl}^2N^2}\right)\phi^{\pi} + \frac{k^2-3\,K}{a^2}\,\psi^{\pi} + \frac{3\,H}{N}\,\dot{\psi}^{\pi} + \frac{k^2(k^2-3\,K)}{6\,a^2}\, E^{\pi} 
\right.
\nonumber\\
&&\left.\qquad\qquad\qquad-\frac{V'}{2\,M_{Pl}^2}\,\delta\sigma^{\pi} -\frac{\dot{\sigma}}{2\,M_{Pl}^2N^2}\delta\dot{\sigma}^{\pi}\right]\,,
\label{nondyn}
\end{eqnarray}
where
\begin{equation}
 \delta\sigma^{\pi} \equiv \delta\sigma - \dot{\sigma}\pi^0, 
\end{equation}
and substitute their solutions into the action. Defining the analogue of
the Sasaki-Mukhanov variable 
\begin{equation}
Q\equiv \delta\sigma^{\pi}-\frac{\dot{\sigma}}{N\,H}\,\left(\psi^{\pi} +\frac{k^2}{6}\,E^{\pi}\right),
\label{SasMuk}
\end{equation}
the resulting quadratic action reads, up to boundary terms,
\begin{equation}
I^{(2)}_{scalar}=\frac{M_{Pl}^2}{2} \int d^3k\,dt\, a^3\,N
 \,\left[6\,M_{GW}^2|\psi^{\pi}|^2-\frac{1}{6}\,M_{GW}^2k^2\left(k^2-3\,K\right)| E^{\pi}|^2 + {\cal L}_Q\right]\,,
\end{equation}
where
\begin{equation}
{\cal L}_Q\equiv\frac{2\, H^2(k^2-3\,K)}{2\, H^2M_{Pl}^2\,(k^2-3\,K)+K\,\frac{\dot{\sigma}^2}{N^2}}\left(\frac{1}{N^2}\,|\dot{Q}|^2-M_Q^2\,|Q|^2\right),
\label{Qact}
\end{equation}
and
\begin{equation}
M^2_Q \equiv V'' + \frac{k^2}{a^2}\left(1-\frac{a^2\,\dot{\sigma}^2}{K M_{Pl}^2\,N^2}\right)+\frac{2\,\left(k^2\, H \,\frac{\dot{\sigma}}{N}+K\,V'\right)\,\left[\left(k^2\, H^2-\frac{K^2}{a^2}\right)\frac{\dot{\sigma}}{N}+K\, H\,V'\right]
}{K\,M_{Pl}^2\, H\left(2\, H^2\,(k^2-3\,K)+K\frac{\dot{\sigma}^2}{M_{Pl}^2N^2}\right)}\,.
\end{equation}
Provided that the parameters ($\alpha_3$, $\alpha_4$) take generic
values away from the special values shown in (\ref{MGWzero}), the
equation of motion for $\psi^{\pi}$ and $E^{\pi}$ lead to 
$\psi^{\pi}=E^{\pi}=0$ and then
\begin{equation}
I^{(2)}_{scalar}=\frac{M_{Pl}^2}{2} \int d^3k\,dt\, a^3\,N
 {\cal L}_Q\,. \label{Qac}
\end{equation}
As shown in Appendix~\ref{app:comparison}, the action (\ref{Qac}) agrees
with the standard results in GR coupled to the same
scalar matter field $\sigma$.

To summarize, the scalar sector consists of a dynamical degree which
evolves exactly like the standard Sasaki-Mukhanov variable in GR, and
two more degrees which have infinite mass.

\section{Summary and discussions}

In the context of the potentially ghost-free, nonlinear massive gravity
\cite{deRham:2010kj} and for a general fiducial metric of FRW form, we
analyzed linear perturbations around self-accelerating cosmological
solutions of arbitrary spatial curvature, populated by generic matter  
content that is minimally coupled to gravity. By constructing a gauge
invariant formulation of perturbations, we found that massive
graviton modes in the scalar and vector sectors have vanishing kinetic
terms but nonzero mass terms. By integrating them out, we showed 
that the part of the action quadratic in scalar and vector perturbations
is exactly the same as in GR with the same matter content. In other
words, the dynamical degrees in the gravity sector comprise only the two
gravity wave polarizations. We also found that these acquire
a mass whose time dependence is set by the fiducial metric.

In Fierz-Pauli theory in de Sitter background, it has been known that
the scalar mode among five degrees of freedom of massive spin-$2$ 
graviton becomes ghost for $2H^2>m_{FP}^2$, where $H$ is the Hubble
expansion rate and $m_{FP}$ is the graviton
mass~\cite{Higuchi:1986py}. This conclusion does not hold in the 
nontrivial cosmological branches of the nonlinear massive
gravity. Indeed, as stated above, the scalar and vector modes have
vanishing kinetic terms and nonzero mass terms for any FRW
background. This sharp contrast to the linear (Fierz-Pauli) massive
gravity stems from a peculiar structure of the graviton mass term
expanded up to the quadratic order in perturbations: it depends only on
the ($ij$)-components of metric perturbations and thus are independent
of ($00$) and ($0i$)-components. This Lorentz-violating structure is
possible because the vev of $f_{\mu\nu}$ in the cosmological branches
spontaneously breaks diffeomorphism invariance in a nontrivial way.


It is still fair to say that the nature of the cancellation of the
kinetic terms in the quadratic action of scalar and vector sectors is
not well understood. This may be an indication that these sectors
exhibit an infinitely strong coupling. If this is the case then we
cannot properly describe the scalar and vector sectors without knowledge
of a UV completion. On the other hand, since the modes have non
vanishing masses, it may be possible that these are just infinitely
heavy modes without low-energy dynamics, as we have assumed in the main
part of the present paper. In this case we can safely integrate out
those extra modes and trust the resulting low energy effective theory.
In order to establish the fate of those extra degrees of freedom, i.e to
judge whether they are strongly-coupled or non-dynamical, the linear
perturbation theory is not sufficient and nonlinear methods are needed.
This study is beyond the scope of the present paper and is left for a
future work.

On the other hand, the modification in the tensor sector may leave a
signature in the stochastic gravitational wave spectrum. The additional
term in the mass of the tensor modes is time dependent, while its sign
is determined by both the fiducial metric and the cosmological
evolution. A positive but large contribution may give rise to a
suppression of the gravity waves and a null signal in the large scale
tensor-to-scalar ratio. However, this deviation from scale invariance
may allow the signal at small scales to be potentially observable in the
space-based gravity wave observatories such as DECIGO \cite{DECIGO}, BBO
\cite{BBO} and LISA \cite{Danzmann:1997hm}.


It is important to note that the present analysis is purely classical
and special care is needed when discussing the evolution of cosmological
perturbations which start off in quantum mechanical vacuum. In order to
address quantum issues such as the radiative stability of the structure
of the effective theory describing cosmological perturbations, one of the
first important steps is to identify the strong coupling scale below
which the nonlinear massive gravity theory can be trusted. In the branch
described by the trivial solution to the St\"uckelberg equation of
motion (\ref{eqn:constraintf}), the so called decoupling limit has been useful for this purpose.
However, since this trivial branch is not compatible with FRW
cosmologies, in the present paper we have considered other two branches
described by non-trivial solutions (\ref{eq:fsolcosmo}). 
In these cosmological branches, the usual decoupling limit is
not applicable, at least apparently. We thus need to develop a new
technique to identify the strong coupling scale, or directly analyze
nonlinear dynamics of the whole system. This is certainly one of
the most important issues in the future research.

\begin{acknowledgments}
 The authors thank G.~D'Amico, C.~de Rham, N.~Kaloper, K.~Koyama,
 N.~Tanahashi and A.~J.~Tolley for useful discussions. This work was
 supported by the World Premier International Research Center Initiative
 (WPI Initiative), MEXT, Japan. S.M. also acknowledges the support by
 Grant-in-Aid for Scientific Research 17740134, 19GS0219, 21111006,
 21540278, by Japan-Russia Research Cooperative Program. 
\end{acknowledgments}

\appendix

\section{Calculation of graviton mass term}
\label{app:gravitonmass}

In this appendix we consider the graviton mass term introduced in
Sec.~\ref{setup} and generalized at the beginning of
Sec.~\ref{sec:perturbations}. We thus consider arbitrary $K$,
$n(\varphi^0)$ and $\alpha(\varphi^0)$. Without loss of generality, we
assume that $N> 0$, $n> 0$, $a>0$ and
$\alpha>0$ at least in the vicinity of the time of interest, where $N$
and $a$ are the lapse function and the scale factor for the background
FRW physical metric.

We expand the generalized graviton mass term up to the quadratic order
in perturbations. In doing so, we shall not use the background equations
of motion for the physical metric since they depend not only on the
graviton mass term but also on the Einstein-Hilbert action and the
matter action. On the other hand, we shall use the background equation
of motion for the St\"uckelberg fields in the middle of calculation.

\subsection{Background equation of motion for St\"uckelberg fields}

In this subsection we shall derive the background equation of motion for
the St\"uckelberg fields $\varphi^a$ by expanding the graviton mass term
$I_{mass}$ up to the linear order in $\pi^a$ ($=\delta\varphi^a$)
without variation of the physical metric $g_{\mu\nu}$. This will be a
good warm-up for the forthcoming subsections.

Using the formula (\ref{eqn:fmunu-expansion}), $f_{\mu\nu}$ is expanded
up to the linear order as
\begin{eqnarray}
 f_{00} & = & 
  -n^2\left[ 1 + \frac{2}{n}\partial_t(n\pi^0) + O(\epsilon^2) \right],
  \nonumber\\
 f_{0i} & = & f_{i0}  =  
  \alpha n \left[ - \frac{n}{\alpha}D_i\pi^0
	    + \frac{\alpha}{n}\dot{\pi}_i + O(\epsilon^2) \right], \nonumber\\
 f_{ij} & = & \alpha^2
  \left[(1+2nH_f\pi^0)\Omega_{ij} 
   + D_i\pi_j+D_j\pi_i + O(\epsilon^2) \right],
\end{eqnarray}
where an overdot represents differentiation with respect to the time
$t$ and
\begin{equation}
 H_f \equiv \frac{\dot{\alpha}}{n\alpha}.
  \label{eqn:def-Hf}
\end{equation}
With the unperturbed physical metric
\begin{equation}
 g_{00}=-N^2, \quad g_{0i}=g_{i0}=0, \quad g_{ij}=a^2\Omega_{ij},
\end{equation}
this leads to the following expansion for $f^{\mu}_{\ \nu}$ 
($\equiv g^{\mu\rho}f_{\rho\nu}$).
\begin{eqnarray}
 f^0_{\ 0} & = & 
  \frac{n^2}{N^2}
  \left[ 1 + \frac{2}{n}\partial_t(n\pi^0) + O(\epsilon^2) \right],
  \nonumber\\
 f^0_{\ i} & = & 
  -\frac{\alpha n}{N^2} \left[ - \frac{n}{\alpha}D_i\pi^0
	    + \frac{\alpha}{n}\dot{\pi}_i + O(\epsilon^2) \right], \nonumber\\
 f^i_{\ 0} & = & 
  \frac{\alpha n}{a^2} \left[ - \frac{n}{\alpha}D^i\pi^0
	    + \frac{\alpha}{n}\dot{\pi}^i + O(\epsilon^2) \right], \nonumber\\
 f^i_{\ j} & = & \frac{\alpha^2}{a^2}
  \left[(1+2nH_f\pi^0)\delta^i_j + D^i\pi_j+D_j\pi^i + O(\epsilon^2) \right].
\end{eqnarray}

Then, using the formula in Appendix~\ref{app:square-root} for matrix
square-root, ${\cal K}^{\mu}_{\ \nu}$ defined by
(\ref{Kdef})-(\ref{eqn:square-root}) is expanded up to the linear order
as 
\begin{equation}
 {\cal K}^{\mu}_{\ \nu} = {\cal K}^{(0)\mu}_{\quad\ \nu} + 
{\cal K}^{(1)\mu}_{\quad\ \nu} + O(\epsilon^2),
\end{equation}
where
\begin{equation}
{\cal K}^{(0)0}_{\quad\ 0} = 1 - \frac{n}{N}, \quad
{\cal K}^{(0)0}_{\quad\ i} = 0, \quad 
{\cal K}^{(0)i}_{\quad\ 0} = 0, \quad
{\cal K}^{(0)i}_{\quad\ j} = \left(1-\frac{\alpha}{a}\right)\delta^i_{\ j}, 
\end{equation} 
and
\begin{eqnarray}
{\cal K}^{(1)0}_{\quad\ 0} &= &
 -\frac{1}{N}\partial_t(n\pi^0), \nonumber\\
{\cal K}^{(1)0}_{\quad\ i} & = & 
 \frac{na}{N^2(1+r)}\left[ - \frac{n}{\alpha}D_i\pi^0
 + \frac{\alpha}{n}\dot{\pi}_i\right], 
\nonumber\\
{\cal K}^{(1)i}_{\quad\ 0} & = & 
 -\frac{n}{a(1+r)}\left[ - \frac{n}{\alpha}D^i\pi^0
	    + \frac{\alpha}{n}\dot{\pi}^i\right], \nonumber\\
{\cal K}^{(1)i}_{\quad\ j} & = & 
 -\frac{\alpha}{2a}
  \left[2nH_f\pi^0\delta^i_j + D^i\pi_j+D_j\pi^i\right].
\end{eqnarray}
Here, we have defined
\begin{equation}
 r \equiv \frac{na}{N\alpha}.
\end{equation}

It is now straightforward to expand the graviton mass term
(\ref{eqn:Imass}) up to the first order. The result is
\begin{equation}
  I_{mass} = I^{(0)}_{mass}  
   + M_{Pl}^2m_g^2
   \int d^4x Na^3\sqrt{\Omega}
   \frac{3n}{a}(aH-\alpha H_f)J_{\phi}\pi^0
   + O(\epsilon^2), 
\end{equation}
where the zero-th order part $I^{(0)}_{mass}$ does not depend on $\pi^a$
and 
\begin{equation}
 J_{\phi} \equiv 3-2X+\alpha_3(1-X)(3-X)+\alpha_4(1-X)^2, \quad
  X \equiv \frac{\alpha}{a}. 
  \label{eqn:def-Jf}
\end{equation}

Therefore, the background equation of motion for the St\"uckelberg
fields is 
\begin{equation}
(aH-\alpha H_f)J_{\phi} = 0,
\end{equation}
where $H_f$ is defined in (\ref{eqn:def-Hf}) and 
\begin{equation}
 H \equiv \frac{\dot{a}}{Na}. 
\end{equation}
Setting $aH=\alpha H_f$ would not allow nontrivial cosmologies since in
this case the background evolution of the physical metric would be
determined not by the matter content but by the fiducial metric. Thus,
we restrict our attention to solutions of $J_{\phi}=0$. This leads to
$X=X_{\pm}$, where $X_{\pm}$ are given by (\ref{eq:fsolcosmo}).

\subsection{Unitary gauge}

In this subsection we shall expand the mass term up to the quadratic
order in the unitary gauge, i.e. under the gauge condition
\begin{equation}
 \pi^0 = \pi^i = 0. 
\end{equation}
After obtaining the expression in the unitary gauge, it is relatively
easy to infer the corresponding expression in general gauge and to
confirm it. The expression in general gauge will be presented in the
next subsection.

In the unitary gauge, the fiducial metric $f_{\mu\nu}$ is the same as in
the background. Hence, $f^{\mu}_{\ \nu}$ 
($\equiv g^{\mu\rho}f_{\rho\nu}$) is expanded up to the second order as 
\begin{eqnarray}
 f^0_{\ 0} & = & \frac{n^2}{N^2}
  \left[1-2\phi+(4\phi^2-\beta^i\beta_i)+O(\epsilon^3)\right], \nonumber\\
 f^0_{\ i} & = & \frac{\alpha^2}{Na}
  \left[\beta_i-2\phi\beta_i-\beta^jh_{ji}+O(\epsilon^3)\right], \nonumber\\
 f^i_{\ 0} & = & -\frac{n^2}{Na}
  \left[\beta^i-2\phi\beta^i-\beta_jh^{ji}+O(\epsilon^3)\right], \nonumber\\
 f^i_{\ j} & = & \frac{\alpha^2}{a^2}
  \left[\delta^i_{\ j}-h^i_{\ j}-\beta^i\beta_j+h^{ik}h_{kj}+O(\epsilon^3)\right],
\end{eqnarray}
where
\begin{equation}
 \beta^i \equiv \Omega^{ij}\beta_j, \quad
  h^i_{\ j} = h_j^{\ i} \equiv \Omega^{ik}h_{kj}, \quad
  h^{ij} \equiv \Omega^{ik}\Omega^{jl}h_{kl}. 
\end{equation}
Then, taking the matrix square-root as prescribed in Appendix~\ref{app:square-root}, 
${\cal K}^{\mu}_{\ \nu}$ defined by (\ref{Kdef})-(\ref{eqn:square-root})
is expanded up to the quadratic order as
\begin{equation}
 {\cal K}^{\mu}_{\ \nu} = {\cal K}^{(0)\mu}_{\quad\ \nu} + 
{\cal K}^{(1)\mu}_{\quad\ \nu} + {\cal K}^{(2)\mu}_{\quad\ \nu} 
+ O(\epsilon^3),
\end{equation}
where
\begin{equation}
{\cal K}^{(0)0}_{\quad\ 0} = 1 - \frac{n}{N}, \quad
{\cal K}^{(0)0}_{\quad\ i} = 0, \quad 
{\cal K}^{(0)i}_{\quad\ 0} = 0, \quad
{\cal K}^{(0)i}_{\quad\ j} = \left(1-\frac{\alpha}{a}\right)\delta^i_{\ j}, 
\end{equation} 
\begin{equation}
{\cal K}^{(1)0}_{\quad\ 0} = \frac{n}{N}\phi, \quad
{\cal K}^{(1)0}_{\quad\ i} = -\frac{\alpha\beta_i}{N(1+r)},\quad
{\cal K}^{(1)i}_{\quad\ 0} = \frac{nr\beta^i}{a(1+r)},\quad
{\cal K}^{(1)i}_{\quad\ j} = \frac{\alpha}{2a}h^i_{\ j},
\end{equation}
and
\begin{eqnarray}
{\cal K}^{(2)0}_{\quad\ 0} & = & \frac{n}{N}
  \left[ -\frac{3}{2}\phi^2 + \frac{r(2+r)}{2(1+r)^2}\beta^k\beta_k
  \right], \nonumber\\
{\cal K}^{(2)0}_{\quad\ i} & = & \frac{\alpha}{N(1+r)}
  \left[ \frac{2+r}{1+r}\phi\beta_i + \frac{1+2r}{2(1+r)}\beta^kh_{ki}
  \right], \nonumber\\
{\cal K}^{(2)i}_{\quad\ 0} & = & -\frac{nr}{a(1+r)}
  \left[ \frac{2+r}{1+r}\phi\beta^i + \frac{1+2r}{2(1+r)}\beta_kh^{ki}
  \right], \nonumber\\
{\cal K}^{(2)i}_{\quad\ j} & = & \frac{\alpha}{a}
  \left[ \frac{1+2r}{2(1+r)^2}\beta^i\beta_j - \frac{3}{8}h^{ik}h_{kj}
       \right].
\end{eqnarray}

It is then straightforward to expand $\left[{\cal K}^n\right]$
($n=1,2,3,4$) up to the quadratic order. The result is
\begin{equation}
 \left[{\cal K}^n\right] = \left[{\cal K}^n\right]^{(0)} 
  + \left[{\cal K}^n\right]^{(1)} + \left[{\cal K}^n\right]^{(2)} + O(\epsilon^3), 
\end{equation}
where
\begin{eqnarray}
 \left[{\cal K}^n\right]^{(0)} & = & 3(1-X)^n+(1-rX)^n, \nonumber\\
 \left[{\cal K}^n\right]^{(1)} & = & 
  nrX(1-rX)^{n-1}\phi + \frac{n}{2}X(1-X)^{n-1}h,
\end{eqnarray}
and 
\begin{eqnarray}
 \left[{\cal K}\right]^{(2)} & = & -\frac{3}{2}rX\phi^2 
  -\frac{3}{8}Xh^{ij}h_{ij} + \frac{r_2X}{2r_1}\beta^i\beta_i, 
  \nonumber\\
 \left[{\cal K}^2\right]^{(2)} & = & (4rX-3)rX\phi^2
  +\left(X-\frac{3}{4}\right)Xh^{ij}h_{ij}
  + \frac{X}{r_1}(r_2-r_3 X)\beta^i\beta_i, \nonumber\\
 \left[{\cal K}^3\right]^{(2)} & = & 
  - \frac{3}{2}(3-5rX)(1-rX)rX\phi^2
  -\frac{3}{8}(3-5X)(1-X)Xh^{ij}h_{ij}\nonumber\\
 & &   + \frac{3X}{2r_1}(r_2-2r_3X+r_4X^2)\beta^i\beta_i ,
  \nonumber\\
 \left[{\cal K}^4\right]^{(2)} & = & 
  6(2rX-1)(1-rX)^2rX\phi^2 + \frac{3}{2}(2X-1)(1-X)^2Xh^{ij}h_{ij}
  \nonumber\\
 & &+ \frac{2X}{r_1}(r_2-3r_3X+3r_4X^2-r_5X^3)\beta^i\beta_i. 
\end{eqnarray}
Here, $X$  is defined in (\ref{eqn:def-Jf}) and 
\begin{equation}
  r_n \equiv \sum_{i=0}^{n}r^n.
\end{equation}

We are now ready to expand the graviton mass term (\ref{eqn:Imass}) up
to the quadratic order. However, before doing so, let us expand it up to
the linear order in order to see the effective energy density due to the
graviton mass term. The result is
\begin{equation}
 I_{mass} = I^{(0)}_{mass} + 
  \int dx^4Na^3\sqrt{\Omega}
  \left[-\left(\phi+\frac{1}{2}h\right)\rho_g
   +\frac{1}{2}M_{Pl}^2m_g^2(1-r)XhJ_{\phi}\right] + O(\epsilon^2),
  \label{eqn:unitary-linear}
\end{equation}
where the zero-th order part $I^{(0)}_{mass}$ does not depend on
perturbations, and 
\begin{equation}
 \rho_g = -M_{Pl}^2m_g^2(1-X)
  \left[3(2-X)+\alpha_3(1-X)(4-X)+\alpha_4(1-X)^2\right] 
\end{equation}
is the effective energy density due to the graviton mass.

Having obtained the expression for $\rho_g$ and noticing that the factor 
$(\phi+h/2)$ in (\ref{eqn:unitary-linear}) is the linear order part of 
\begin{equation}
 \frac{\sqrt{-g}}{Na^3\sqrt{\Omega}} = 
  1 + \left(\phi+\frac{1}{2}h\right)
  + \left[ - \frac{1}{2}\phi^2 + \frac{1}{2}\beta^i\beta_i
     + \frac{1}{8}(h^2-2h^{ij}h_{ij})+ \frac{1}{2}\phi h
    \right] + O(\epsilon^3),
\end{equation}
we expect that expanding 
\begin{equation}
 \tilde{I}_{mass}[g_{\mu\nu},f_{\mu\nu}] \equiv 
  I_{mass}[g_{\mu\nu},f_{\mu\nu}] + 
  \int d^4x\sqrt{-g}\rho_g, \label{eqn:def-Itildemass-offshell}
\end{equation}
instead of $I_{mass}$ itself, should simplify the resulting
expression. We thus expand $\tilde{I}_{mass}$ up to the quadratic
order. The result is 
\begin{equation}
 \tilde{I}_{mass} = 
  M_{Pl}^2m_g^2\int d^4x Na^3\sqrt{\Omega}
  \left[\tilde{L}_{mass}^{(0)}+\tilde{L}_{mass}^{(1)}
   +\tilde{L}_{mass}^{(2)}+O(\epsilon^3)\right],
\end{equation}
where
\begin{eqnarray}
 \tilde{L}_{mass}^{(0)} & = & 
  -rX(1-X)\left[3+3\alpha_3(1-X)+\alpha_4(1-X)^2\right], \nonumber\\
 \tilde{L}_{mass}^{(1)} & = & \frac{1}{2}(1-r)Xh J_{\phi}, \nonumber\\
 \tilde{L}_{mass}^{(2)} & = & \frac{1}{2}
  \left[\phi h + \frac{\beta^i\beta_i}{1+r}
   +\frac{1}{4}(1-r)(h^2-2h^{ij}h_{ij})\right]XJ_{\phi} \nonumber\\
 & & 
  + \frac{1}{8}m_g^{-2}M_{GW}^2(h^2-h^{ij}h_{ij}),
\end{eqnarray}
and 
\begin{equation}
 m_g^{-2}M_{GW}^2 = 
  XJ_{\phi} + (1-r)X^2\left[1+\alpha_3(2-X)+\alpha_4(1-X)\right].
\end{equation}

As shown in the previous subsection, the background equation of motion
for the St\"uckelberg fields for nontrivial cosmological branches is
$J_{\phi}=0$ and gives $X=X_{\pm}$, where $J_{\phi}$ and 
$X_{\pm}$ are defined in (\ref{eqn:def-Jf}) and
(\ref{eq:fsolcosmo}), respectively. For $X=X_{\pm}$, it is easy to see that
$\rho_g=M_{Pl}^2\Lambda_{\pm}$, where $\Lambda_{\pm}$ are 
defined in (\ref{lambdapm}). Thus, upon using $J_{\phi}=0$, the
definition of $\tilde{I}_{mass}$ in (\ref{eqn:def-Itildemass-offshell})
reduces to (\ref{eqn:def-Itildemass}). Also, for $X=X_{\pm}$, $M_{GW}^2$
defined above reduces to that defined in (\ref{eqn:MGW2}).

In summary, in the unitary gauge, upon using the background equation of
motion $J_{\phi}=0$ for the St\"uckelberg fields but without using the
background equation of motion for the physical metric,
$\tilde{I}_{mass}$ defined in (\ref{eqn:def-Itildemass}) is expanded up
to the quadratic order as 
\begin{equation}
 \tilde{I}_{mass}[g_{\mu\nu},f_{\mu\nu}] = 
 \tilde{I}^{(0)}_{mass}
 + \frac{M_{Pl}^2} {8}\int d^4x Na^3\sqrt{\Omega}
 M_{GW}^2(h^2-h^{ij}h_{ij}) + O(\epsilon^3),
\end{equation}
where the zero-th order part
$\tilde{I}_{mass}^{(0)}\equiv
M_{Pl}^2m_g^2\int d^4x Na^3\sqrt{\Omega}\tilde{L}^{(0)}_{mass}$
does not depend on the perturbations.

\subsection{General gauge}

In general gauge, we expect that the expansion of $\tilde{I}_{mass}$
defined in (\ref{eqn:def-Itildemass-offshell}) should be similar to that
in the unitary gauge, provided that each metric perturbation variable is
replaced by the corresponding gauge-invariant variable constructed from
the metric perturbation and the St\"uckelberg field perturbation. Such
gauge-invariant variables are defined in
(\ref{eqn:phipi-betapi-hpi}). We thus expect that 
\begin{equation}
 \tilde{I}_{mass}[g_{\mu\nu},f_{\mu\nu}] = 
 \tilde{I}^{(0)}_{mass}
 + \frac{M_{Pl}^2} {8}\int dx^4 Na^3\sqrt{\Omega}
 M_{GW}^2\left[(h^{\pi})^2-h_{\pi}^{ij}h^{\pi}_{ij}\right]
 + \Delta_{mass}, 
 \label{eqn:tildeImass-expansion-general}
\end{equation}
with relatively simple expression for $\Delta_{mass}$ up to the
quadratic order. Of course, for $J_{\phi}=\pi^a=0$, $\Delta_{mass}$
should vanish up to the quadratic order, as shown in the previous
subsection. Indeed, by direct computation we can confirm that 
this is true not only in unitary gauge $\pi^a=0$, but also for
$\pi^a\ne 0$ as far as $J_{\phi}=0$ is imposed: 
\begin{equation}
 \Delta_{mass} = O(\epsilon^3) \,,\quad  \mbox{for}\quad
  J_{\phi}=0. \label{eqn:Deltamass-onshell}
\end{equation}

If we do not impose $J_{\phi}=0$ nor $\pi^a=0$ then $\Delta_{mass}$ 
up to the quadratic order is 
\begin{eqnarray}
 \Delta_{mass} & = & M_{Pl}^2m_g^2\int dx^4Na^3\sqrt{\Omega}
  \left\{
   \left[ \frac{1-r}{2}Xh^{\pi}-3\dot{X}\pi^0\right]\right.\nonumber\\
 & & 
  + \frac{1}{2}\left[\phi^{\pi}h^{\pi}
		+\frac{\beta_{\pi}^i\beta^{\pi}_i}{1+r}
		+\frac{1}{4}(1-r)(h^2-2h^{ij}h_{ij})
	       \right.\nonumber\\
 & & 
  -(1-r)(\dot{h}+3NHh)\pi^0-\frac{2N}{a}(1-r)\beta^iD_i\pi^0\nonumber\\
 & & 
  -\frac{a^2}{N^2(1+r)}D^i\dot{\pi}^0D_i\dot{\pi}^0
  - \frac{N^2}{a^2}(1-r)D^i\pi^0D_i\pi^0
 + \frac{a^2}{N^2(1+r)}\dot{\pi}^i\dot{\pi}_i
  \nonumber\\
& & \left.
     + 3(1-r)\pi^0
     \left(\frac{1}{a^3}\partial_t(Na^3H\pi^0)
      +NHD_i\pi^i\right)
    \right]X\nonumber\\
& & 
 \left.
      +\frac{3}{2}
      \left[-(2\phi+h)+\frac{1}{Na^3}\partial_t(Na^3\pi^0) +
       D_i\pi^i\right]\pi^0\dot{X} 
       \right\}J_{\phi}
 + O(\epsilon^3).
\end{eqnarray}

\section{Quadratic action for the scalar field example}
\label{app:acttot}

This appendix presents the total quadratic action for the scalar field
example of Sec.~\ref{sec:example-scalar}. Combining the graviton mass
term
(\ref{eqn:tildeImass-expansion-general})-(\ref{eqn:Deltamass-onshell}) 
with the Einstein-Hilbert and scalar field parts, the complete quadratic
action can be calculated up to boundary terms as 
\begin{eqnarray}
I^{(2)} &=& M_{Pl}^2 \int d^4 x \,N\,a^3\,\sqrt{\Omega}\,\Bigg\{
{\cal L}+\frac{\sqrt{-g}^{(2)}}{Na^3\sqrt{\Omega}}\left[3\,H^2+\frac{3\,K}{a^2}-\Lambda_\pm-\frac{1}{M_{Pl}^2}\,\left(\frac{\dot{\sigma}^2}{2\,N^2}+V\right)\right]\nonumber\\
&&\qquad\qquad\qquad\qquad\qquad\qquad
+\frac{1}{8}\left(h^2-2\,h_{ij}h^{ij}\right)\left(\frac{2\,\dot{H}}{N}-\frac{2\,K}{a^2}+\frac{\dot{\sigma}^2}{M_{Pl}^2N^2}\right)
\Bigg\}\,,
\label{totact}
\end{eqnarray}
where
\begin{equation}
\frac{\sqrt{-g}^{(2)}}{Na^3\sqrt{\Omega}} = -\frac{\phi^2}{2}+\frac{1}{2}\beta_i\beta^i-\frac{1}{4}h_{ij}h^{ij}+\frac{1}{8}\,h^2+\frac{1}{2}\,\phi \,h\,,
\end{equation}
and
\begin{eqnarray}
 {\cal L} &=& 
\frac{1}{8\,N^2}\left(\dot{h}_{ij}\dot{h}^{ij}-\dot{h}^2\right)+\frac{H}{N}\phi\,\dot{h} - \frac{1}{a}\left(2\,H\,\phi-\frac{1}{2\,N}\,\dot{h}\right)D_i\beta^i -\frac{1}{2\,N\,a}\,D_i\beta_j\,\dot{h}^{ij}-3\,H^2\,\phi^2
\nonumber\\
&&
+\frac{1}{4\,a^2}\left[D_i\beta_jD^i\beta^j-(D_i\beta^i)^2-2\,K\,\beta^i\beta_i\right] + \frac{1}{2\,a^2}\,\left(D_iD_jh^{ij}-\triangle h\right)\phi
\nonumber\\
&&
+\frac{1}{8\,a^2}\left[2\,D^ih_{ik}\,D_jh^{jk}-D_kh_{ij}\,D^kh^{ij}+2\,h\,D_iD_jh^{ij}-h\,\triangle h\right]
-\frac{K}{4\,a^2}\,\left(h_{ij}h^{ij}+4\,h\,\phi\right)
\nonumber\\
&&+{\cal M}_{GW}^2 \left[\frac{1}{8}\left(h^2- h_{ij}h^{ij}\right)-\frac{1}{2}\,\left(D_i\pi^i+2\,N\,H\,\pi^0\right)h +\frac{1}{2}\,h_{ij}\,D^i\pi^j +3\,H^2\,N^2\,(\pi^0)^2
\right.
\nonumber\\
&&\left.\qquad\qquad\qquad\qquad+\frac{1}{4}\,\left[(D_i\pi^i)^2-D_i\pi_j D^i\pi^j+2\,K\,\pi_i\pi^i\right]+2\,H\,N\,\pi^0D_i\pi^i
\right]
\nonumber\\
&&
+\frac{1}{M_{Pl}^2}\,\left[\frac{\delta\dot{\sigma}^2}{2\,N^2}+\frac{\dot{\sigma}}{2\,N^2}\,\left(h-2\,\phi\right)\delta\dot{\sigma} - \frac{1}{2\,a^2}\,D_i\delta\sigma\,D^i\delta\sigma -\frac{\dot{\sigma}}{a\,N}\,\beta_iD^i\delta\sigma - \frac{V''}{2}\,\delta\sigma^2
\right.
\nonumber\\
&&\qquad\qquad\qquad\qquad\qquad\qquad\qquad\qquad\qquad\qquad
\left.
-\frac{V'}{2}\,\left(h+2\,\phi\right)\delta\sigma +\frac{\dot{\sigma}^2}{2\,N^2}\,\phi^2\right]\,.
\label{totlag}
\end{eqnarray}

\section{Comparison with GR}
\label{app:comparison}

In this appendix we compare the scalar action (\ref{Qac}) with the
GR results in the literature. For this purpose it is
useful to define the gauge invariant Bardeen potential, which in our
language corresponds to 
\begin{equation}
\Phi = \phi^{\pi} +\frac{1}{N}\,\frac{d}{dt}\left(a \beta^{\pi} - \frac{a^2}{2\,N}\,\dot{E}^{\pi}\right)\,,
\end{equation}
or, using Eqs.(\ref{bgeqs}), (\ref{nondyn}) and (\ref{SasMuk}), as well as the equation of motion for Q obtained from varying (\ref{Qact}), we can write
\begin{equation}
\Phi = -\frac{a^2\,H^2}{2\,M_{Pl}^2\,H^2\,(k^2-3K)+K\frac{\dot{\sigma}^2}{N^2}}\,\frac{\dot{\sigma}}{N}
\left[\frac{\dot{Q}}{N}+\left(3\,H - \frac{\dot{\sigma}^2}{2\,M_{Pl}^2\,H\,N^2}+ \frac{N\,V'}{\dot{\sigma}}\right)Q\right]\,.
\end{equation}
Using the above definition, along with the equation of motion for $Q$, then switching to conformal time $ a\,d\eta = N\,dt$,
we see that the equation of motion for $\Phi$ in the standard
scenario, given in Eq.(8.140) of Ref.~\cite{Peter:2009zzc}, is
satisfied. To compare the normalization of the action with the
literature, it is also useful to define the following quantities 
\begin{equation}
\xi \equiv \frac{2\,a\,M_{Pl}^2}{H}\,\Phi \,,
\quad 
\zeta \equiv \frac{N\,H}{\dot{\sigma}}\,Q - \frac{K\,H\,N^2}{a^2\,\dot{\sigma}^2}\,\xi\,.
\end{equation}
With these definitions, it is straightforward to verify that the equations of motion
given in Eqs.(21-22) of Ref.~\cite{Garriga:1999vw} are satisfied, with
substitution $dt \to N dt$. Furthermore, using the above expressions,
the action given in Eq.~(24) of \cite{Garriga:1999vw} reproduces our
action (\ref{Qac}) up to boundary terms.

\section{Perturbative expansion of matrix square root}
\label{app:square-root}

Let us expand an $N\times N$ matrix $A$ as
\begin{equation}
 A = A^{(0)} + \epsilon A^{(1)} + \epsilon^2 A^{(2)} +O(\epsilon^3),
\end{equation}
where the zero-th order part is assumed to be of the form
\begin{equation}
 (A^{(0)})^t_{\ t} = \alpha, \quad (A^{(0)})^t_{\ j} = 0, \quad
  (A^{(0)})^i_{\ t} = 0, \quad (A^{(0)})^i_{\ j} = \beta\delta^i_j, 
\end{equation}
with $\alpha>0$ and $\beta>0$. The square-root is expanded as
\begin{equation}
 \sqrt{A} = B^{(0)} + \epsilon B^{(1)} + \epsilon^2 B^{(2)} + O(\epsilon^3),
\end{equation}
where
\begin{equation}
 (B^{(0)})^t_{\ t} = \sqrt{\alpha}, \quad (B^{(0)})^t_{\ j} = 0,\quad
  (B^{(0)})^i_{\ t} = 0, \quad (B^{(0)})^i_{\ j} = \sqrt{\beta}\delta^i_j,
\end{equation}
\begin{eqnarray}
 (B^{(1)})^t_{\ t} & = & \frac{1}{2\sqrt{\alpha}}(A^{(1)})^t_{\ t}, \quad
  (B^{(1)})^t_{\ j} = \frac{1}{\sqrt{\alpha}+\sqrt{\beta}}(A^{(1)})^t_{\ j}, \nonumber\\
  (B^{(1)})^i_{\ t} & = & \frac{1}{\sqrt{\alpha}+\sqrt{\beta}}(A^{(1)})^i_{\ t}, \quad
  (B^{(1)})^i_{\ j} = \frac{1}{2\sqrt{\beta}}(A^{(1)})^i_{\ j}, 
\end{eqnarray}
and
\begin{eqnarray}
 (B^{(2)})^t_{\ t} & = & \frac{1}{2\sqrt{\alpha}}(\tilde{A}^{(2)})^t_{\ t}, \quad
  (B^{(2)})^t_{\ j} = \frac{1}{\sqrt{\alpha}+\sqrt{\beta}}(\tilde{A}^{(2)})^t_{\ j},  \nonumber\\
  (B^{(2)})^i_{\ t} & = & \frac{1}{\sqrt{\alpha}+\sqrt{\beta}}(\tilde{A}^{(2)})^i_{\ t}, \quad
  (B^{(2)})^i_{\ j} = \frac{1}{2\sqrt{\beta}}(\tilde{A}^{(2)})^i_{\ j}. 
\end{eqnarray} 
Here, $\tilde{A}^{(2)}\equiv A^{(2)} -B^{(1)\,2}$.

\end{document}